\documentclass[twocolumn,prb,amsmath,amssymb]{revtex4}

\usepackage{graphics}
\usepackage{graphicx}
\usepackage{dcolumn}
\usepackage{bm}

\begin{document}

\title{Statistical properties of charged interfaces}

\author{S.~Teber}

\affiliation{Laboratoire de Physique Th\'eorique et Mod\`eles Statistiques,
B\^atiment 100, Universit\'e Paris-Sud, 91406, Orsay-C\'edex, France}

\begin{abstract}

 We consider the equilibrium statistical properties of interfaces submitted
to competing interactions; a long-range repulsive Coulomb interaction inherent
to the charged interface and a short-range, anisotropic, attractive one due
to either elasticity or confinement. We focus on one-dimensional interfaces
such as strings. Model systems considered for applications are mainly aggregates
of solitons in polyacetylene and other charge density wave systems, domain lines
in uniaxial ferroelectrics and the stripe phase of oxides. At zero temperature,
we find a shape instability which lead, via phase transitions, to tilted phases.
Depending on the regime, elastic or confinement, the order of the zero-temperature
transition changes. Thermal fluctuations lead to a pure Coulomb roughening of
the string, in addition to the usual one, and to the presence of angular kinks.
We suggest that such instabilities might explain the tilting of stripes in cuprate
oxides. The \( 3D \) problem of the charged wall is also analyzed. The latter
experiences instabilities towards various tilted phases separated by a tricritical
point in the elastic regime. In the confinement regime, the increase of dimensionality
favors either the melting of the wall into a Wigner crystal of its constituent
charges or a strongly inclined wall which might have been observed in nickelate
oxides.

\end{abstract}
\maketitle

\section{Introduction}

Various types of systems display peculiar properties due to competing long-range
forces. In this paper we focus on the statistical properties of charged interfaces,
in Ising-like systems, such as aggregates of charged topological defects in
charge-density wave systems \cite{SBNK,Yu Lu} and among them polyacetylene
\cite{Heeger}, charged domain-lines in uniaxial ferroelectrics \cite{Monceau}
or stripes in oxides \cite{Tranq,Zaanen}. Instabilities, such as the
observed inclination of stripes in cuprates \cite{incli} and manganese \cite{Dai},
might be related to the presence of the long range Coulomb interactions and
its competition with an attractive force. Uniaxial ferroelectrics and density
waves are also model systems where such instabilities could take place. The
present study deals with one or two-dimensional charged interfaces, i.e. strings
or walls. The former is realized in oxides, monolayers of doped conducting polymers
or other charge-density waves and junctions in field effect experiments \cite{battlogg}
for equivalent materials. The latter might be present, due to a dimensional
crossover which induces a \( 3D \) ordering, in the systems cited above.

\noindent The above interests emerged from a model-independent theory concerning
the statistical and thermodynamic properties of uncharged and charged solitons which has
been studied, respectively, in Refs.~\onlinecite{bohr} and~\onlinecite{teber}. The system considered
is two-fold degenerate and, when topologically doped, the solitons form a one-component
plasma in either \( 2 \) or \( 3D \) space. The competition between the long-range
\( 3D \) Coulomb interaction of the particles and the confinement force between
them has led to a very rich phase diagram. Particularly interesting was the
regime where the temperature is much less than the confinement energy scale,
so that the solitons are actually bound into pairs, and the repulsive Coulomb
interaction is weak enough to preserve these bisolitons as the elementary particles.
In this regime, aggregated phases of bisolitons were shown to exist. The study
of such aggregates in the continuum limit is equivalent to the study of charged
interfaces.

\noindent The statistical properties of charged fluctuating manifolds have been
considered before \cite{mehran}. In this study, the authors have been considering
a generalization of the theory of polyelectrolytes, cf. Ref.~\onlinecite{orland}, and
have been focusing on the properties emerging from scale invariance and renormalization.
Even though some of our results will converge with the previous study, which
seems to be quite natural, our aim will be different. We consider interfaces
directed along an anisotropy axis of the system which are either in an elastic
or confinement regime; the latter is relevant to quasi-one dimensional systems
in relation with the statistical properties of solitons mentioned above. In both regimes
the Coulomb interaction favors the disintegration of the interface. In contrast
with the instability towards a modulation which has been suggested in Ref.~\onlinecite{teber}
this competition, as we will show, leads to \textit{zero-temperature shape instabilities}
with respect to a \textit{tilting} of the interface. We will focus mainly on
the properties related to this discrete symmetry breaking. This will able us
to better understand the structure of the solitonic lattice in the presence
of the long range repulsive interaction. In this respect, and as explained in
\cite{teber}, even though this problem seems to be quite old, the detailed
impact of the Coulomb interaction has not been reported before. Moreover, the
charged solitonic lattice is commonly observed nowadays as stripes in various
oxides. Experimentally the observation that upon doping the system, at low temperatures,
a transition from a collinear, i.e. vertical, to a diagonal stripe phase takes
place \cite{incli} is strikingly similar to our actual results.

\noindent The paper is organized as follows. We will mainly concentrate on the
statistical properties of a charged string. In section \ref{presentation} we present the
models of the confined and elastic charged strings. In section \ref{zerotemp}
we derive the statistical properties of the string with the help of a saddle-point calculation.
We present the zero-temperature results dealing with the instabilities of the
string. In section \ref{finitetemp} the effects of thermal fluctuations are
considered. In section \ref{numerics} the previous results are confirmed with
the help of a numerical approach. In section \ref{applications} the present
study is applied to the \( 2D \) and \( 3D \) solitonic lattices, as well
as stripes and uniaxial ferroelectrics.

\section{Elastic and confinement regimes for the charged string}

\label{presentation}The problem of the charged string \cite{teber} has emerged
from the study of the statistical properties of topological defects, i.e. solitons, common
to charge density wave (CDW) systems. These general arguments are reproduced
in the following as they give a full meaning to the physical origins of the
present theory, cf. \cite{SBNK} for a review on the physics of topological
defects in CDW systems. It should however be clear that this study is also relevant
to other systems with discrete symmetry breaking as has been mentioned already
in the Introduction.

\bigskip{}
\noindent For the moment we consider a one-dimensional system, i.e. a single
chain, in a CDW state. The latter is described by a lattice order parameter
which is complex in general \[\Delta (x)=|\Delta (x)|\exp (i\varphi (x))\]
where \( x \) runs along the chain, \( |\Delta (x)| \) is related to the CDW
gap and \( \varphi (x) \) fixes the position of the CDW with respect to a host
lattice. In weakly commensurate CDW the low energy excitations are excitations
of the phase. This leads, cf. \cite{schultz}, to a sine-Gordon type Hamiltonian
\begin{equation}
\label{SineGordon}
H=\int dx\left( \frac{\hbar v_{F}}{2\pi }\left( \frac{\partial \varphi }{\partial x}\right) ^{2}-W\cos \left( M\varphi \right) \right)
\end{equation}
where the first term gives the elastic energy of the CDW and the second term
reflects the pinning of the CDW superstructure by the host lattice; \( M \)
is the degeneracy of the ground-state.

\noindent The non-trivial solutions of the equation of motion associated with
(\ref{SineGordon}) are the solitons, i.e. the \( \varphi - \)particles \cite{rice},
which may be viewed as compressions or dilatations of the CDW. As we are in
the realm of electronic crystals the solitons might therefore carry a charge
which is fractional, i.e. \( q=2/M \), in the general case. For a double-degenerate
ground state, \( M=2 \), they are given by
\begin{equation}
\label{phiSG}
\varphi (x)=2\arctan \left( \exp (2x/\xi )\right)
\end{equation}
where the length \( \xi \sim \sqrt{\hbar v_{F}/W} \) corresponds to the width
of this nucleus domain wall connecting the two ground states \( \varphi (\pm \infty )=\pm 1 \).

\noindent The situation is slightly different for the two-fold degenerate trans-polyacetylene
because in this case the band is half-filled and the system is commensurate.
In particular equation (\ref{SineGordon}) does not describe this case. However
the previous arguments remain qualitatively the same. The defect is an amplitude
or \( \pi - \)soliton; it carries an integer charge and is described by
\begin{equation}
\label{SSHsoliton}
\Delta (x)=\Delta _{0}\tanh (x/\xi )
\end{equation}
where \( \Delta _{0} \) is a constant deformation of the chains. Again this
excitation connects the two ground states of the system \( \Delta (\pm \infty )=\pm \Delta _{0} \).

\bigskip{}
\noindent For a quasi-one-dimensional system, which we consider here, neighboring
chains must have the same phase modulo \( 2\pi  \). This leads to the following
general two-particle interaction
\begin{equation}
\label{InterchainInt}
E(\delta x)=-V\int dz\Delta _{\alpha }(\delta x-z)\Delta _{\beta }(z)
\end{equation}
between neighboring chains \( \alpha  \) and \( \beta  \), with coupling constant
\( V \), and solitons distant by \( \delta x \). In the coarse-grained regime,
\( \delta x\gg \xi  \), (\ref{InterchainInt}) grows linearly, \( E\approx V\delta x \),
a signature of the confinement of the defects. This can be shown explicitly
with the help of (\ref{SSHsoliton}) and (\ref{InterchainInt}). Dealing with
a string, e.g. a domain line of solitons, this confinement energy is equivalently
reproduced by the Solid on Solid (SOS) model. The related Hamiltonian reads
\begin{equation}
\label{SOS}
H_{conf}=\frac{J_{\perp }}{a_{y}}\sum ^{L-1}_{y=0}|x_{y+1}-x_{y}|
\end{equation}
where we have taken \( J_{\perp }=Va_{y} \) as the inter-chain energy scale,
\( y \) runs along the vertical direction and \( x_{y} \) is an integer variable
giving the deviation of the string with respect to the \( y \) axis.

\noindent The opposite case, corresponding to small deviations \( \delta x\ll \xi  \),
is more general for the energy then increases quadratically \( E\approx V\delta x^{2} \).
This is equivalent to the elastic model which is extensively used in interface
physics. The related Hamiltonian reads
\begin{equation}
\label{Gaussian}
H_{elast}=J_{\perp }\int \frac{dy}{a_{y}}\left( \frac{dx}{dy}\right) ^{2}.
\end{equation}

\noindent These two interface models are well known \cite{forgacs}. To compactify
notations we shall define
\begin{equation}
\label{GalInterfaceH}
H_{0p}=J_{\perp }\int \frac{dy}{a^{p}_{y}}|\frac{dx}{dy}|^{p}
\end{equation}
where \( p=1,2 \) leading respectively to the confinement and elastic regimes
and where the continuum limit has been taken in both cases. We thus see that
a crossover (corresponding to fractional \( 1<p<2 \)) from the elastic to the
confinement regime takes place as the distance between solitons increases, i.e.
as the tilt angle of the string increases. This is manifested by the fact that
\( \delta x=(dx/dy)a_{y}=\zeta a_{y} \) where \( \zeta  \) is related to the
tilt angle of the string and \( a_{y} \) is the inter-chain distance. For \( \zeta \ll \xi /a_{y} \)
the string is elastic whereas for \( \zeta \gg \xi /a_{y} \) the confined string
is relevant. The crossover between the two regimes takes place at
\begin{equation}
\label{croosover angle}
\zeta _{0}=\xi /a_{y}.
\end{equation}

\noindent The introduction of the \( 3D \) long-range Coulomb interaction yields
\begin{widetext}
\begin{equation}
\label{HCoulomb}
H_{c}=\frac{(ze)^{2}}{2\epsilon }\int \int dydy'\left( \frac{1}{\sqrt{(y-y')^{2}+(x(y)-x(y'))^{2}}}-\frac{1}{\sqrt{(y-y')^{2}}}\right)
\end{equation}
\end{widetext}
where \( z\leq 1 \) so that the elementary constituent of the string might
carry a fractional charge \( ze \) (\( z=1 \) for polyacetylene), \( \epsilon  \)
is the dielectric constant of the isotropic, neutral media in which the plane
is embedded and the contribution of the vertical string has been subtracted.

\noindent The thermodynamics of the charged interface is governed by \( H_{p}=H_{0p}+H_{c} \)
or explicitly
\begin{widetext}
\begin{equation}
\label{TotHamExplicit}
H_{p}=J_{\perp }\int \frac{dy}{a^{p}_{y}}|\frac{dx}{dy}|^{p}+\frac{(ze)^{2}}{2\epsilon }\int \int dydy'\left( \frac{1}{\sqrt{(y-y')^{2}+(x(y)-x(y'))^{2}}}-\frac{1}{|y-y'|}\right) .
\end{equation}
\end{widetext}

\section{The charged string at zero temperature}

\label{zerotemp}

\subsection{The saddle point approximation}

We consider a configuration
\begin{equation}
\label{Distr}
x(y)=x^{(0)}(y)+\delta x(y)
\end{equation}
were \( x(y) \) has been expanded in the vicinity of the saddle point distribution
\( x^{(0)}(y) \) with deviations \( \delta x(y) \). Free boundary conditions
are taken. The expanded Hamiltonian reads, in the harmonic approximation
\begin{widetext}
\begin{equation}
\label{Ham4e}
H_{p}=H_{p}^{(0)}+\frac{1}{2}\int \int dydy'\delta x(y)\delta ^{(2)}H/\delta x(y)\delta x(y')|_{\delta x=0}\delta x(y')+...
\end{equation}
\end{widetext}
where \( H_{p} \) is given by (\ref{TotHamExplicit}) and \( H_{p}^{(0)}\equiv H_{p}\{\delta x^{(0)}(y)\} \).
The saddle point distribution is given by \( \delta H_{p}/\delta x(y)=0 \)
that is
\begin{widetext}
\begin{equation}
\label{EaMotExplicit}
J_{\perp }p\left( \Delta x^{(0)}(y)\right) ^{p-1}=\frac{(ze)^{2}}{2\epsilon }\int ^{+\infty }_{-\infty }dy'\frac{x^{(0)}(y')-x^{(0)}(y)}{\left( \left( y-y'\right) ^{2}+\left( x^{(0)}(y)-x^{(0)}(y')\right) ^{2}\right) ^{3/2}}.
\end{equation}
\end{widetext}
This non-linear integro-differential equation cannot be solved exactly. We
will therefore take an ansatz for the ground-state configuration \( x^{(0)}(y) \)
of the string. As can be seen from equation (\ref{TotHamExplicit}), at zero
temperature, the elastic or confinement term favors a vertical string, i.e.
along the \( y \) axis, whereas the Coulomb favors the disintegration of the
string in the perpendicular direction. Instabilities of the string arise from
the competition between these interactions. \textit{We find that these instabilities
lead to a tilted string as the new ground state of the system}. Therefore
\begin{equation}
\label{DistTilted0}
x^{(0)}(y)=\zeta y
\end{equation}
where \( \zeta =\tan \theta  \), \( \theta  \) being the tilt angle of the
string with respect to the \( y \) axis. This ansatz satisfies the linearized
equation (\ref{EaMotExplicit}) up to some logarithmic corrections. As will
be shown explicitly in the following, \( \zeta  \) is related to the ratio
between the Coulomb and confinement or elastic energy scales. It is thus related
to the elementary charge of the string constituent soliton.

\bigskip{}
\noindent With the help of (\ref{DistTilted0}), and (\ref{TotHamExplicit}),
(\ref{Ham4e}) reads
\begin{equation}
\label{HamWithTiltedDist}
H_{p}=H_{p}^{(0)}+\sum _{k}\lambda ^{2}_{p,k}(\zeta _{p})|\delta x_{k}|^{2}
\end{equation}
with the eigenvalues
\noindent
\begin{equation}
\label{omega}
\lambda ^{2}_{p,k}(\zeta _{p})=k^{2}J_{\perp }\left( p(p-1)+\frac{2\zeta _{p}^{2}-1}{\left[ 1+\zeta _{p}^{2}\right] ^{\frac{5}{2}}}\gamma (k)\right) .
\end{equation}
\( \zeta _{p} \) is the optimal angle corresponding to the distribution (\ref{DistTilted0})
minimizing \( H^{(0)}_{p} \) and
\begin{equation}
\label{gamma}
\gamma (k)=\gamma \log (1/ka)
\end{equation}
with
\begin{equation}
\label{gamma0}
\gamma =\frac{w_{y}}{J_{\perp }}\qquad w_{y}=\frac{(ze)^{2}}{2\epsilon a_{y}},
\end{equation}
\( \gamma  \) being the ratio of the Coulomb energy scale \( w_{y} \) to the
elastic or confinement energy scales.

\noindent The first term in (\ref{HamWithTiltedDist}) corresponds to the mean-field
Hamiltonian. \( H_{p}^{(0)} \) leads to the following mean-field free energy
density
\begin{equation}
\label{MFfreeEn}
f_{p}^{(0)}(\zeta )=\frac{J_{\perp }}{a_{y}}\left[ \zeta ^{p}+\gamma _{D}\left( \frac{1}{\sqrt{1+\zeta ^{2}}}-1\right) \right]
\end{equation}
where \( \gamma _{D}=\gamma (l^{-1}_{D}) \) and \( l_{D} \) is the Debye screening
length taken into account as a hard cut-off to eliminate the logarithmic divergence
of the Coulomb term in the thermodynamic limit. Details concerning the screening
mechanisms and the expressions of \( l_{D} \) have been reported in Appendix
A. As can be seen in the first term of (\ref{MFfreeEn}) in the confinement
regime, the absolute value has been dropped. This is the non-return approximation
for the confinement regime the limitations of which will be considered at the
end of this section. In the following subsections we will see how the tilted
phase arises via zero-temperature phase transitions in both \( p=1,2 \) regimes.

\subsection{The elastic charged string}

\begin{figure}
\includegraphics[width=8cm,height=5cm]{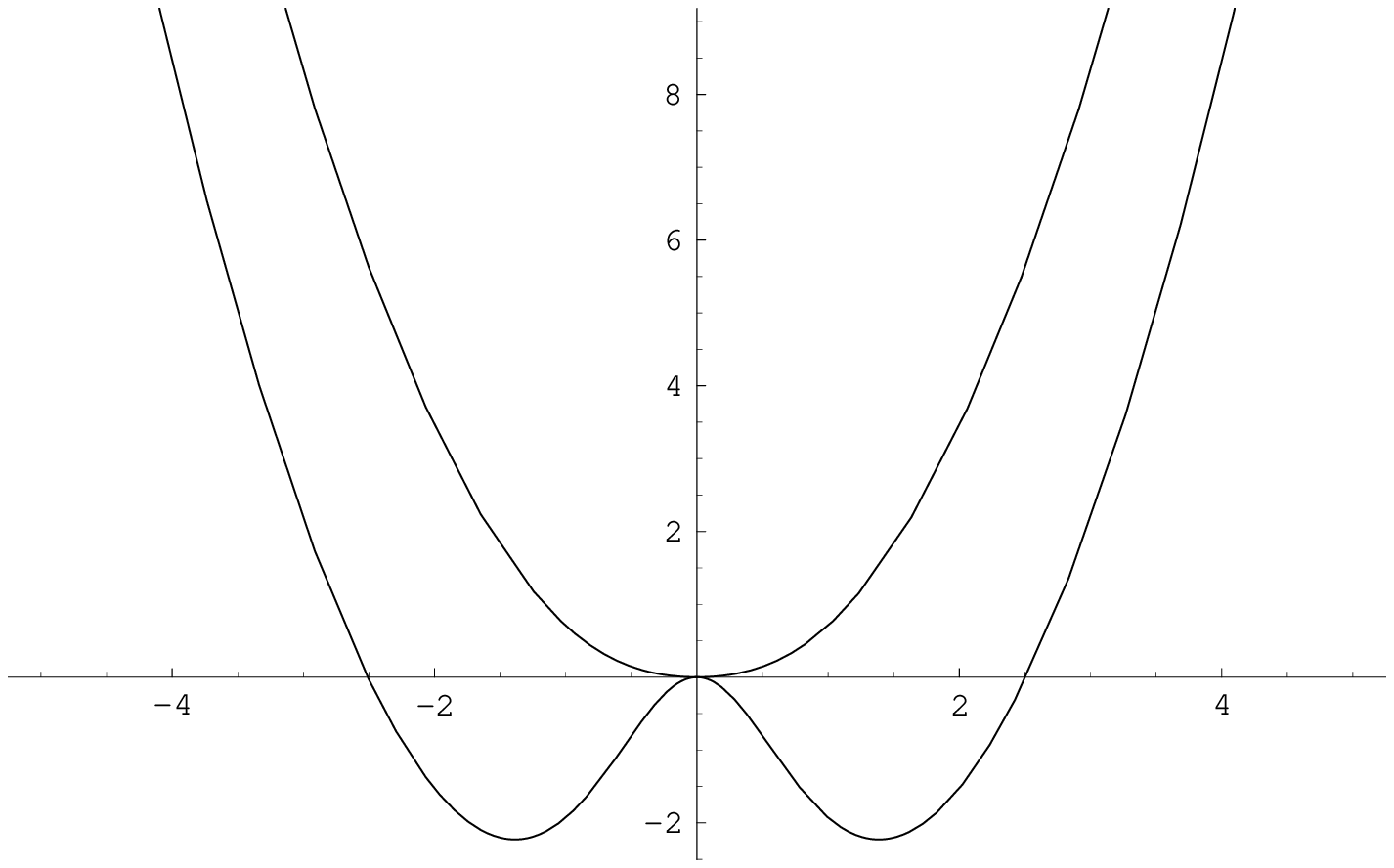}
\includegraphics[width=8cm,height=5cm]{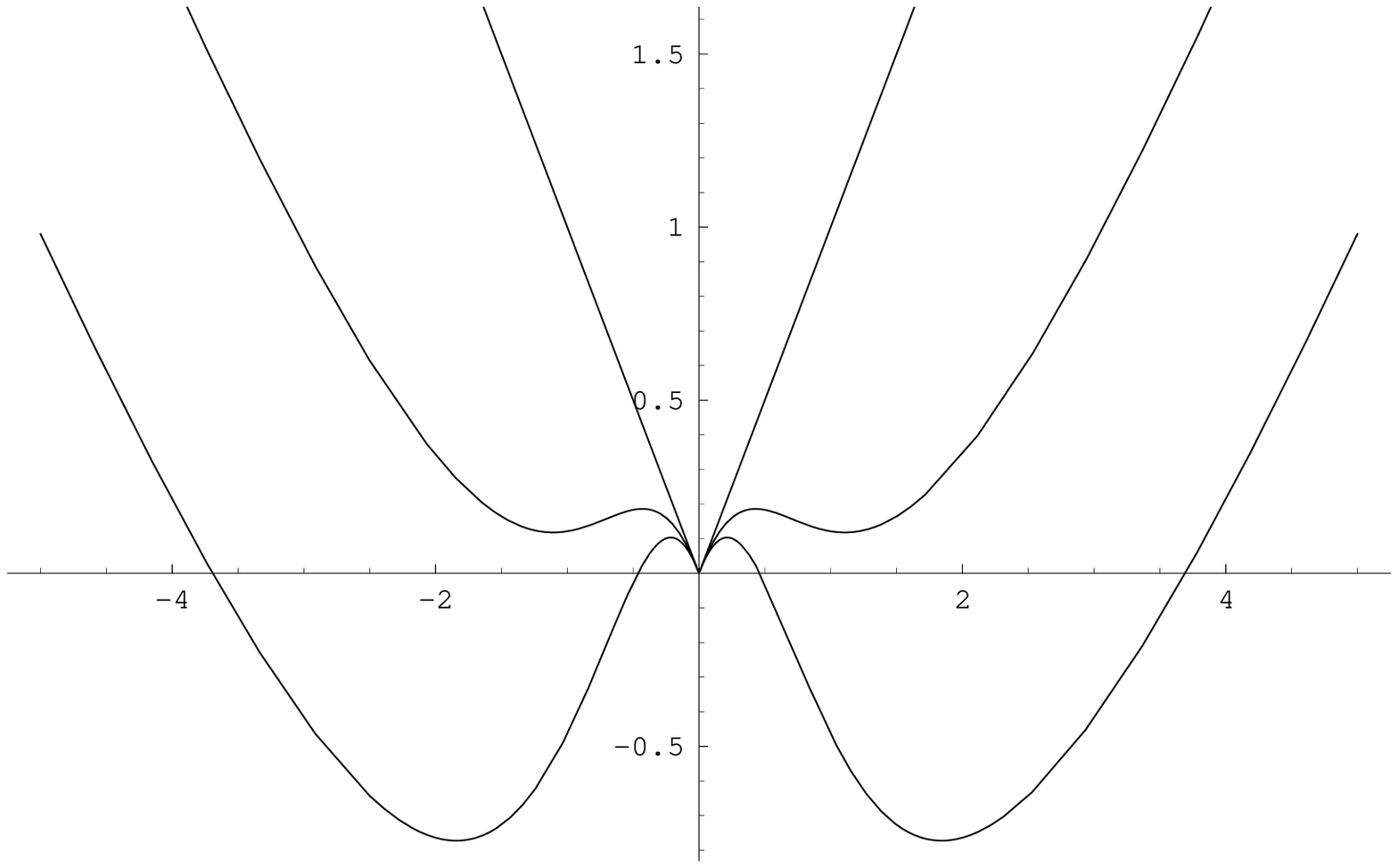}
\caption{  \label{Figure 1} 1) Mean-field free energy of the elastic string (\ref{MFfreeEn})
as a function of \protect\( \zeta \protect \) exhibiting a second order phase
transition. The two curves correspond to \protect\( \gamma _{D}<\gamma ^{c}_{D}\protect \)
where the vertical string is stable and \protect\( \gamma _{D}>\gamma ^{c}_{D}\protect \)
where the tilted line is stable as can be seen from the double well.
2) Free energy of the confined string as a function of \( \zeta  \)
exhibiting a first order transition from the vertical string to the free string.
From upper to lower curve: free model \( \gamma _{D}=0 \), metastable states
appear when \( \gamma _{D}>\gamma ^{*}_{D} \), they become stable states when
\( \gamma _{D}>\gamma ^{c}_{D} \).}
\end{figure}

\noindent We first consider the elastic regime, with \( p=2 \) in (\ref{MFfreeEn}),
corresponding to tilt angles below the crossover value (\ref{croosover angle}).
Figure \ref{Figure 1} displays this free energy as a function of the angle.
Until
\begin{equation}
\label{gammaCriticGauss}
\gamma ^{c}_{D}=2
\end{equation}
the vertical line, \( \zeta _{2}=0 \), is stable. Above \( \gamma ^{c}_{D} \)
the vertical line becomes unstable and the free energy has a double well shape.
This is due to the double degeneracy of the system; the new, tilted, ground
states corresponding to \( \pm \zeta _{2} \) have the same energies. The optimal
angle reads
\begin{equation}
\label{xiopt2}
\zeta _{2}=\sqrt{\left( \frac{\gamma _{D}}{2}\right) ^{\frac{2}{3}}-1}.
\end{equation}

\noindent The transition at \( \gamma ^{c}_{D} \) is of the second order. This
can be shown by evaluating a quantity analogous to the heat capacitance in usual
thermodynamic phase transitions. That is
\begin{equation}
\label{Cgamma}
C_{\gamma }=-\gamma _{D}\frac{\partial ^{2}f_{2}^{(0)}}{\partial \gamma ^{2}_{D}}
\end{equation}
with the tilt angle given by (\ref{xiopt2}) above the transition and vanishing
below. Eq. (\ref{Cgamma}) yields
\[
C_{\gamma }=\frac{2J_{\perp }}{9}\left( \frac{2}{\gamma _{D}}\right) ^{2/3}\left[ 1+\frac{1}{2}\left( \frac{\gamma _{D}}{2}\right) ^{1/3}\right] \qquad \gamma _{D}>\gamma ^{c}_{D}\]
and \( C_{\gamma }=0 \) otherwise, which leads to a jump at the transition,
\( \Delta C_{\gamma ^{c}_{D}}=J_{\perp }/3 \).

\noindent The linear stability analysis of these mean-field solutions can be
performed by looking at (\ref{omega}) with \( p=2 \)
\begin{equation}
\label{omegaGauss}
\lambda ^{2}_{2,k}(\zeta _{2})=J_{\perp }k^{2}\left( 2+\frac{2\zeta _{2}^{2}-1}{\left[ 1+\zeta _{2}^{2}\right] ^{\frac{5}{2}}}\gamma (k)\right) .
\end{equation}
These eigenvalues must be positive, for all momentum \( k \), in order to
ensure the stability of the solution. 

\noindent Expanding first around the vertical line, \( \zeta _{2}=0 \), (\ref{omegaGauss})
leads to the following stability criterion
\[
\gamma (k)=\gamma \log (1/ka)<2,\: \forall k\]
 or 
\[
k>k_{c}=\frac{1}{a}\exp (-2/\gamma _{0})\]
where \( k_{c} \) is the Coulomb-dependent critical momentum. This shows that
the instability is driven by low-momentum, i.e. long wavelength, modes. Reminding
that \( 1/l_{D}<k<1/a \), we find, in agreement with the mean-field analysis,
that the vertical string is stable as long as \( k_{c}<1/l_{D} \). \( k_{c}=1/l_{D} \)
returns us to (\ref{gammaCriticGauss}). Increasing the strength of the Coulomb
interaction increases \( k_{c} \). For \( k_{c}>1/l_{D} \) a continuous set
of modes, i.e. between \( 1/l_{D} \) and \( k_{c} \), has negative eigenvalues
so that the vertical string is no more stable.

\noindent Expanding around \( \xi _{2}\neq 0 \), we see from (\ref{omegaGauss}),
that all modes are stable as soon as \( \xi _{2}>1/\sqrt{2} \). For such tilt
angles we are of course above the critical point as can be seen with the help
of (\ref{xiopt2}). The tilted line is thus stable against harmonic fluctuations.

\subsection{The charged string in the confinement regime}

\noindent When \( \zeta  \) becomes larger than the crossover value \( \zeta >\zeta _{0} \),
cf. (\ref{croosover angle}), the elastic approximation is no more valid and
we reach the peculiar confinement regime with \( p=1 \). In this case the eigenvalues
(\ref{omega}) are given by
\begin{equation}
\label{omegaSOS}
\lambda ^{2}_{1,k}(\zeta _{1})=J_{\perp }\frac{2\zeta _{1}^{2}-1}{\left[ 1+\zeta _{1}^{2}\right] ^{\frac{5}{2}}}k^{2}\gamma (k).
\end{equation}

\noindent We see from (\ref{omegaSOS}) that configurations with \( \zeta _{1}>1/\sqrt{2}=\zeta ^{*}_{1} \)
are always stable against harmonic fluctuations. Moreover this is a pure Coulomb
stability. On the other hand configurations with \( \zeta _{1}<1/\sqrt{2}=\zeta ^{*}_{1} \)
are always unstable. This implies that the stability of the vertical string
cannot be analyzed within the present model. This is related to the non-return
approximation made in (\ref{HamWithTiltedDist}). Numerical simulations however
show that the tilted string is, here also, the new ground state, cf. section
\( 4 \). In relation with the crossover between elastic and confinement regime
we can then interpret \( \zeta ^{*}_{1} \) as a low boundary for the crossover
angle \( \zeta _{0} \), cf. (\ref{croosover angle}). In the following we will
thus consider solutions corresponding to \( \zeta _{1}>\zeta ^{*}_{1} \), i.e.
\( \gamma _{D}>\gamma ^{*}_{D} \).

\noindent Figure \ref{Figure 1} shows a crucial difference between the confinement
and elastic regimes concerning the mechanism by which tilted strings, i.e. a
double well in the free energy, appear. Contrary to the previous case the second
derivative of the energy remains positive a signature of the fact that the correlation
length remains finite and that the transition is first order in the confinement
regime. Metastable states thus appear. We are going to show first that they
appear at \( \gamma ^{*}_{D} \), the upper spinodal strength. From (\ref{MFfreeEn})
with \( p=1 \), the vanishing of the first derivative of the free energy density
leads to
\begin{equation}
\label{SOSsol}
1=\gamma _{D}\frac{\zeta _{1}}{(1+\zeta ^{2}_{1})^{3/2}}.
\end{equation}

\noindent The spinodal line, above which (metastable) solutions appear can be
defined with the help of the coupled equations (\ref{SOSsol}) and it's derivative.
This leads to
\begin{equation}
\label{metast}
\zeta ^{*}_{1}=\frac{1}{\sqrt{2}}\qquad \qquad \gamma ^{*}_{D}=\frac{3\sqrt{3}}{2}\approx 2.6
\end{equation}
Increasing the Coulomb interaction we reach the critical regime where the mean-field
free energy of the tilted solution becomes equal to that of the \( \zeta =0 \)
solution \( f_{1}(\zeta =0)=f_{1}(\zeta _{1}) \). This equation together with
(\ref{SOSsol}) define the critical point and lead to
\begin{equation}
\label{critical1stOrder}
\zeta ^{c}_{1}=\sqrt{\frac{1}{2}+\frac{\sqrt{5}}{2}}\approx 1.3\qquad \qquad \gamma ^{c}_{D}\approx 3.3
\end{equation}
As in critical phenomena, for \( \gamma _{D}>\gamma ^{c}_{D} \), the system
jumps to the non-zero tilt angle solution corresponding to the absolute minimum
of the energy.

\subsection{Quantum fluctuations}

At \( T=0 \) the string is submitted to quantum fluctuations. The latter give
a contribution to the difference correlation function of the form
\[
<\delta x(y)^{2}>\approx \hbar \log y,\]
where \( y \) is along the string. Assuming that the distance between lines
is of the order of \( l_{D} \), collisions between lines will take place on
a scale \( y\approx \exp l_{D} \) due to these transverse fluctuations. These
collisions are thus present on scales much larger than our upper cut-off \( l_{D} \)
and will therefore not affect previous results.

\section{The charged string at finite temperature}

\label{finitetemp} We seek, in the present section, for the effect of temperature.
In the following we consider the low temperature regime where the string needs
to be quantized. From the expression of the full free energy, thermodynamic
quantities can be evaluated. We give the expression of the heat capacitance.
In the presence of thermal fluctuations, the angular degeneracy gives rise to
defects connecting the ground states of the string: the angular kinks. The \( N- \)kink
problem is defined and the energies of the \( 1- \) and \( 2- \)kink string
configurations are given. Single angular kinks are activated. Bi-angular kinks
are non-topological and are subject to a Coulomb-confinement; these excitations
correspond to a Coulomb-roughening of the string in addition to the usual roughening.

\subsection{Thermodynamic quantities}

\noindent Apart from a logarithmic correction the spectrum (\ref{omega}) is
that of phonons, a feature of the saddle point approximation. We thus have a
harmonic oscillator problem which is easily quantized
\[
E_{\{n_{k}\}}=E^{(0)}+\sum _{k}(n_{k}+\frac{1}{2})\omega _{k,p}\]
 where the units has been chosen such that \( \omega _{k,p} \) has the dimension
of an energy.

\noindent This yields
\begin{equation}
\label{free energy quantized}
f_{p}=f_{p}^{(0)}+T\int ^{\frac{1}{a}}_{\frac{1}{l_{D}}}\frac{dk}{2\pi }\log \left( 1-\exp (-\omega _{k,p}/T)\right)
\end{equation}
where the second term is an entropic contribution and the zero point energy
has been included in the first term. The general expression of the free energy
density is then
\begin{equation}
\label{free energy explicit}
f_{p}=f_{p}^{(0)}-\frac{T^{2}}{12\sqrt{J_{\perp }\left[ p(p-1)+\gamma (\zeta )\right] }}
\end{equation}
where
\[
\gamma (\zeta )=\gamma \frac{2\zeta ^{2}-1}{(1+\zeta ^{2})^{5/2}}\]
 and \( \gamma _{0} \) is given by (\ref{gamma0}).

\noindent It is simple, from (\ref{free energy explicit}), to compute the heat
capacitance, at constant length, of the system. We obtain
\begin{equation}
\label{heat capacitance}
C^{(p)}_{L}=\frac{T}{6\sqrt{J_{\perp }\left[ p(p-1)+\gamma \frac{2\zeta ^{2}-1}{(\zeta ^{2}+1)^{5/2}}\right] }}
\end{equation}
which is linear in \( T \) with an angle dependent coefficient. Such an expression
can be usefully compared with experimental work on related physical systems.

\subsection{Excited states of the string}

\subsubsection{The angular kink solution}

\begin{figure}
\includegraphics[width=8cm,height=5cm]{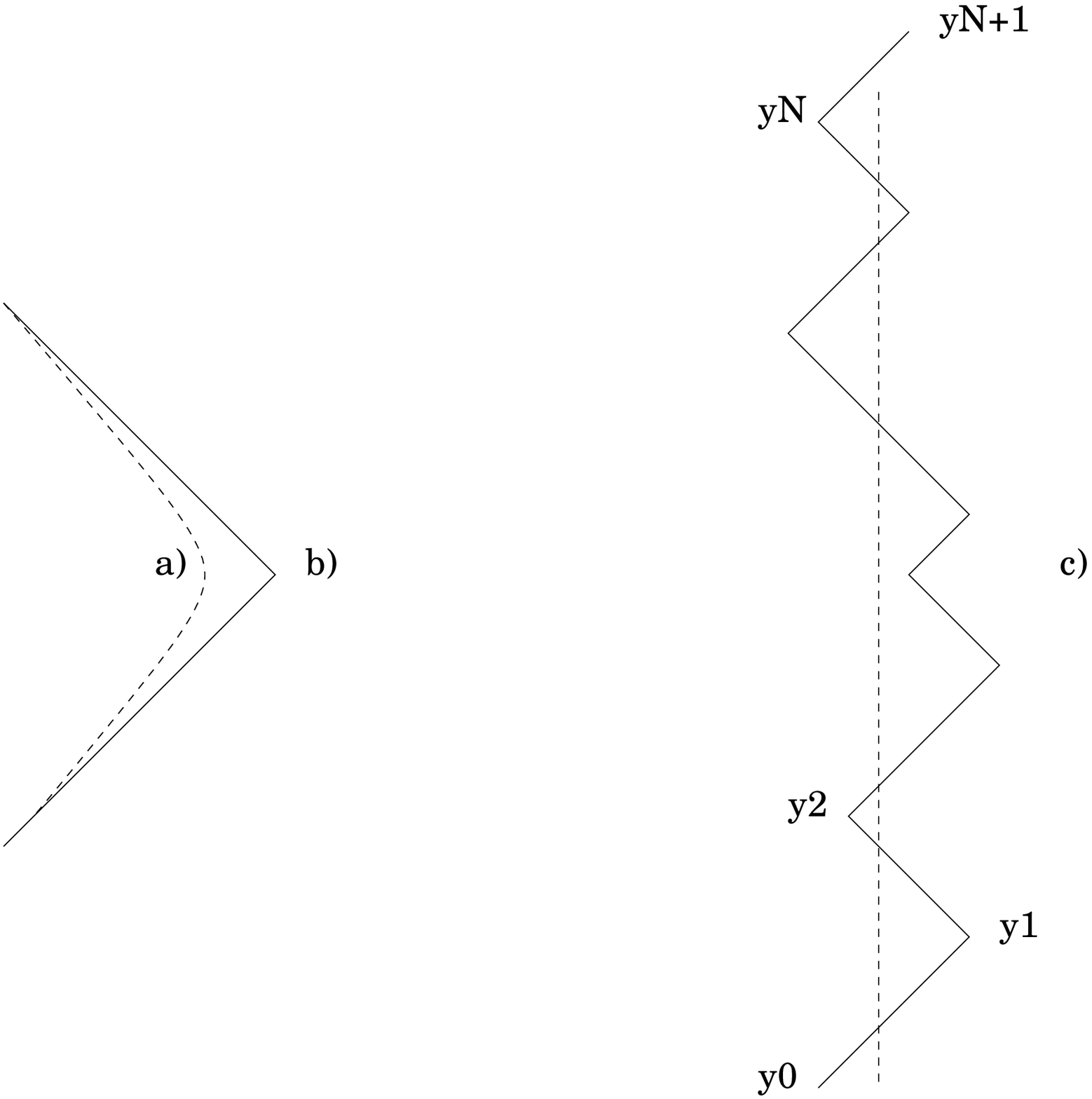}
\includegraphics[width=8cm,height=5cm]{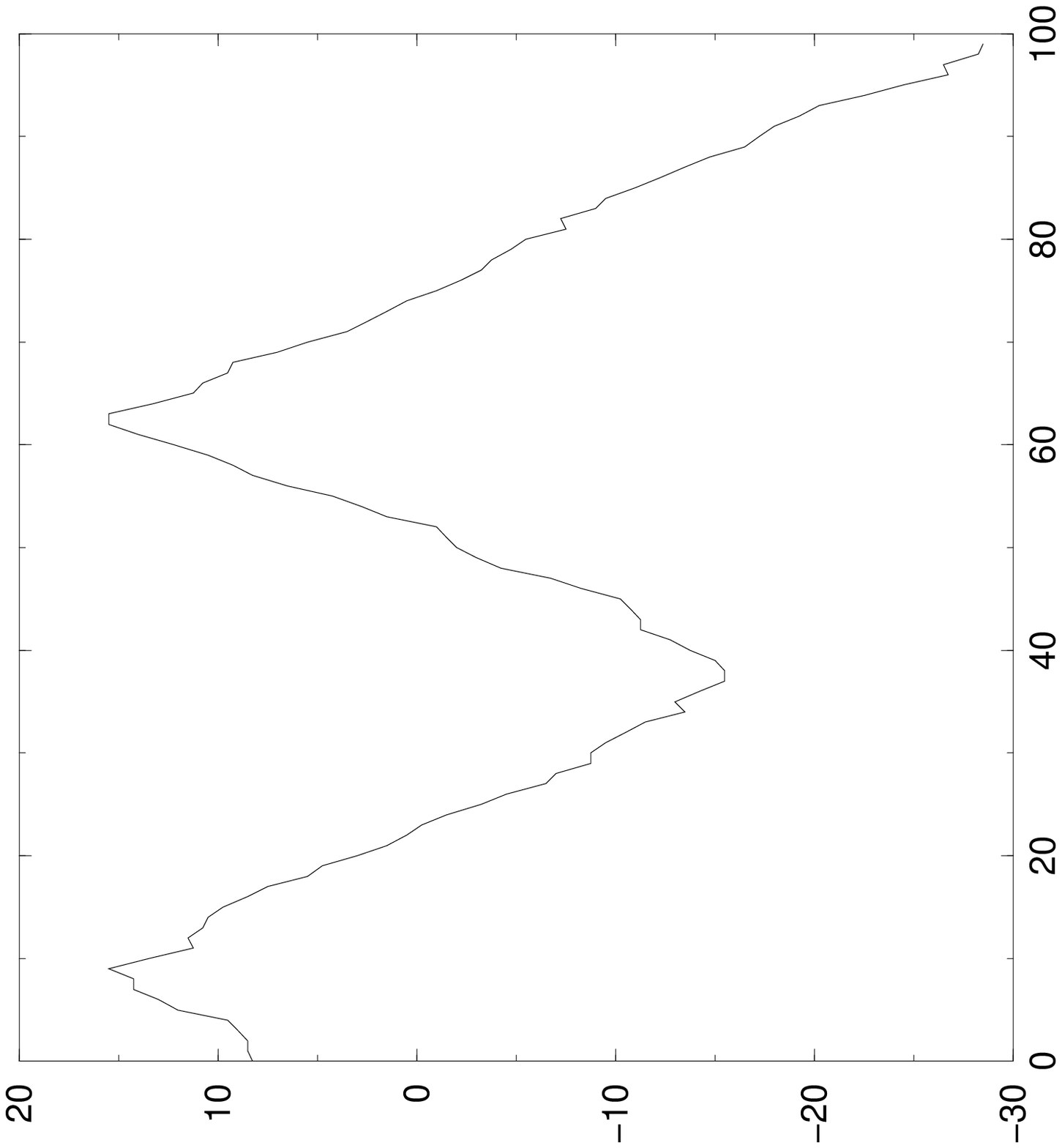}
\caption{ \label{Figure 4}Angular kinks shapes. a) Exact kink shape. b) Point like kink.
c) Shape of the string with \protect\( N-\protect \)kink excitations. d) Numerical
result. }
\end{figure}

\noindent Coarse graining the system we consider the kink as a point-like excitation
as in Figure \ref{Figure 4}b. This amounts to neglect it's core on the scale
of which elastic or confinement energy compete with the Coulomb energy to connect
smoothly the two ground states as shown in Figure \ref{Figure 4}a. At finite
temperatures a general configuration of the string is thus similar to Figure
\ref{Figure 4}c where an array of angular kinks is present. This shape agrees
with the numerical results displayed on Figure \ref{Figure 4}d.

\noindent The energy of an array of \( N \) kinks is given by
\begin{widetext}
\begin{equation}
\label{deltaEN}
\Delta E_{N}=\frac{w_{y}}{a_{y}}\sum ^{N}_{i,j=0}\int ^{y_{i+1}}_{y_{i}}dy\int ^{y_{j+1}'}_{y_{j}'}dy'\frac{1}{\sqrt{(y-y')^{2}+(x_{i+1}(y)-x_{j+1}(y'))^{2}}}-E^{(0)}_{N}
\end{equation}
\end{widetext}
where \( w_{y} \) has been defined in (\ref{gamma0}), \( y_{N+1}=-y_{0}=+\infty  \),
the other \( y_{i} \)'s denoting the position of the defects, \( x_{i}(y) \)
is a portion of string connecting point \( y_{i-1} \) to point \( y_{i} \)
and \( E^{(0)}_{N} \) is the Coulomb energy of the same string without kinks,
i.e. at \( T=0 \). The point-like kink approximation amounts to use the tilted
ansatz ground state \( x_{i}(y)=(-1)^{i}\zeta (y-y_{i})+x_{i}(y_{i}) \). Eq.
(\ref{deltaEN}) is available for both elastic and confinement models. This
is consistent with the fact that these models have similar long distance properties
as shown in Appendix B with the help of a transfer matrix approach. However,
even though exact results can be obtained with the help of (\ref{deltaEN})
within the limits of the point-like kink approximation, the obtained expressions,
in the general case, are quite intractable. We will therefore consider simple
cases.

\noindent First the \( 1- \)kink energy. The latter reads
\begin{equation}
\label{deltaE1}
\Delta E_{1}=\frac{Lw_{y}}{a_{y}\sqrt{1+\zeta ^{2}}}\ln \left[ \frac{1+\sqrt{1+\zeta ^{2}}}{2}\right]
\end{equation}
where \( L \) is the length of the string. The energy of the kink is lower
than the total energy of the string which is over-extensive in the absence of
screening, i.e. \( \sim L\log L \). Nevertheless, Eq. (\ref{deltaE1}) shows
that the kink has infinite energy in the thermodynamic limit, a special feature
of the long range Coulomb interaction. In the presence of screening, the energy
of the string becomes extensive, i.e. \( \sim L\log l_{D} \), and the single
angular kink is activated
\[
\Delta E^{D}_{1}=\frac{l_{D}w_{y}}{a_{y}\sqrt{1+\zeta ^{2}}}\ln \left[ \frac{1+\sqrt{1+\zeta ^{2}}}{2}\right] .\]
 Thus, we have an exponentially small density of kinks
\begin{equation}
\label{DensityOfInstantons}
n\approx \exp \left( -\frac{\Delta E^{D}_{1}}{T}\right) .
\end{equation}

\subsubsection{Transverse fluctuations}

Given the density of kinks (\ref{DensityOfInstantons}), the length of the string
between two angular kinks is given by
\[
l\approx \frac{1}{n}.\]
Within this length thermal fluctuations also play an important role. The difference
correlation function reads, with the help of (\ref{HamWithTiltedDist}), in
the limit of large \( y \)
\begin{equation}
\label{corr func express}
<\left( \delta x(y)-\delta x(0)\right) ^{2}>_{p}\approx y\frac{Ta_{y}}{J_{\perp }}\frac{1}{p(p-1)+\gamma (\zeta )\ln \left( y/a\right) }
\end{equation}
where we have omitted higher order terms. Equation (\ref{corr func express})
shows that the charged string, which might be tilted if the Coulomb interaction
is sufficiently strong, roughens at all non-zero temperatures. This roughening
has two origins: there is a usual roughening due to the elasticity of the string
and a Coulomb roughening, originating from the spontaneous symmetry breaking.
Hence, there is no positional order
\[
<\delta x>\sim \sqrt{<\left( \delta x(y)-\delta x(0)\right) ^{2}>_{p}}\approx \sqrt{y}\rightarrow \infty \qquad y\rightarrow \infty \]
but there is long-range orientational order
\[
<\delta \zeta >\sim \frac{d<\delta x>}{dy}\approx 1/\sqrt{y}\rightarrow 0\qquad y\rightarrow \infty .\]
Notice that the screened Coulomb interaction does not affect the value of the
roughening exponent which is \( 1/2 \). This is also known from a theory of
fluctuating charged manifolds, cf. \cite{mehran}, as has been said in the Introduction,
and confirms the validity of our approach. What is peculiar to our system is
the Coulomb roughening of the string due to the spontaneous symmetry breaking.
The latter might be isolated from the usual roughening by the following procedure.

\noindent We consider the \( 2- \)kink problem on the same line as the single
kink of the previous paragraph. When their separation \( \delta y \), along
the axis of preferable orientation, is smaller than the screening length, equation
(\ref{deltaEN}) yields
\begin{widetext}
\begin{equation}
\Delta E_{2}=\frac{w_{y}}{a_{y}\sqrt{1+\zeta ^{2}}}\sum _{\epsilon =\pm }\epsilon \left\{ \delta y\log
\left( \frac{\sqrt{(1+\zeta ^{2})\left[ (L+\delta y)^{2}+\zeta ^{2}(L-\epsilon \delta y)^{2}\right] }
+L+\delta y+\zeta ^{2}(L-\epsilon \delta y)}{2\delta y\left[ 1+\zeta ^{2}\frac{1-\epsilon }{2}\right] }\right)
\right.
\nonumber
\end{equation}
\begin{equation}
+\delta y\log \left( \frac{\sqrt{(1+\zeta ^{2})\left[ L^{2}+\zeta ^{2}(L-\delta y-\epsilon \delta y)^{2}
\right] }+L+\zeta ^{2}(L-\delta y-\epsilon \delta y)}{2\delta y\left[ 1+\zeta ^{2}\frac{1-\epsilon }{2}
\right] }\right)
\nonumber
\end{equation}
\begin{equation}
+ L\log \left( \frac{\sqrt{(1+\zeta ^{2})\left[ (L+\delta y)^{2}+\zeta ^{2}(L-\epsilon \delta y)^{2}\right] }
+L+\delta y+\zeta ^{2}(\delta y-\epsilon L)}{2L\left[ 1+\zeta ^{2}\frac{1-\epsilon }{2}\right] }\right)
\nonumber
\end{equation}
\begin{equation}
\left. +(L-\delta y)\log \left( \frac{\sqrt{(1+\zeta ^{2})\left[ L^{2}+\zeta ^{2}(L-\delta y-\epsilon
\delta y)^{2}\right] }+L+\zeta ^{2}(\delta y-\epsilon (L-\delta y))}{2(L-\delta y)
\left[ 1+\zeta ^{2}\frac{1-\epsilon }{2}\right] }\right) \right\} .
\label{deltaE2}
\end{equation}
\end{widetext}

\noindent The term with \( \epsilon =+1 \) corresponds to the contribution
of the kinky string whereas the one with \( \epsilon =-1 \) is the subtraction
of the ground-state string. When \( \delta y<l_{D}\ll L \), (\ref{deltaE2})
yields
\begin{equation}
\label{deltaE2small}
\Delta E_{2}\approx \frac{\delta yw_{y}}{a_{y}\sqrt{1+\zeta ^{2}}}\log \left( 1+\zeta ^{2}\right) +\bigcirc \left( (\frac{\delta y}{L})^{2}\right) .
\end{equation}
In the limit \( L\rightarrow \infty  \) the angular bikink has a finite energy.
At the optimal angle, equation (\ref{deltaE2small}) shows that this pair is
\textit{confined} with an energy growing linearly as a function of the transversal
distance between the kinks. The proliferation of angular bikinks is therefore
shown to be a signature of the \textit{Coulomb roughening} of the charged string.

\section{Numerical approach}

\label{numerics}The statistical properties of the charged string in both elastic and confinement
regimes has been studied with the help of a Monte Carlo-Metropolis algorithm.
These numerical simulations allow the computation of the properties of the full
Hamiltonian (\ref{TotHamExplicit}) with open boundary conditions as above.
The ground state is found with the help of a simulated annealing. The shape
of the distribution, the value of the zero temperature order parameter \( \zeta  \)
and the energy of the distribution as a function of \( \gamma  \) have been
computed to detect the various instabilities and associated phase transitions.
The effect of thermal fluctuations, i.e. the roughening and tunneling, have
also been studied. These numerical results show good agreement with the analytical
results presented above.

\subsection{Elastic string}

We focus first on the zero-temperature properties of the string. The ground-state
has been found by performing a logarithmic simulated annealing during each Monte-Carlo
simulation
\[
T[i]=T[0]\frac{\ln 2}{\ln (2+i)}\]
 where \( i \) is the Monte-Carlo step and \( T[0] \) is a high initial temperature.
Due to an improved algorithm to treat the Coulomb interaction the logarithmic
decay of temperature has been successfully implemented in the simulations. As
is well known this leads to the ground-state with the lowest probability of
being trapped in metastable states.

\noindent We compute the properties of the full Hamiltonian (\ref{TotHamExplicit}).
The numerical coupling constant in (\ref{HCoulomb}) is defined by \( \gamma _{num} \).
The rigidity \( J_{\perp } \) in (\ref{Gaussian}) is taken to be unity. With
these numerical coupling constants a comparison can be made with the analytical
results obtained using a mean-field theory. In this respect \( \gamma _{num} \)
should correspond to \( \gamma  \) in (\ref{gamma}). The computational cell
needs of course to be finite and there is therefore no reason for using an upper
cut-off limit \( l_{D} \). Therefore \( l_{D}\rightarrow L \) and \( \gamma _{num}=\gamma _{numD}/\ln L \)
which reflects the non-extensivity arising from the long-range interaction.
In particular the critical numerical coupling constant should be given by \( \gamma ^{c}_{num}=\gamma ^{c}_{numD}/\ln L \)
where \( \gamma ^{c}_{num} \) should be obtained by simulations and \( \gamma ^{c}_{numD} \)
compared with (\ref{gammaCriticGauss}).

\begin{figure}
\includegraphics[width=6.5cm,height=6cm,angle=-90]{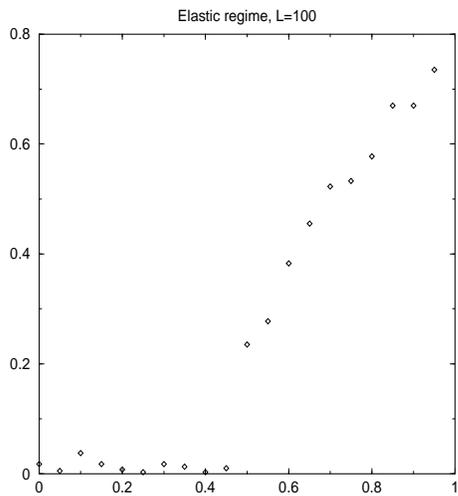}
\caption{\label{GaussXiGamma}Absolute value of \protect\( \zeta \protect \) as a function
of \protect\( \gamma _{num}\protect \)}
\end{figure}

\begin{figure}
\includegraphics[width=8cm,height=5cm]{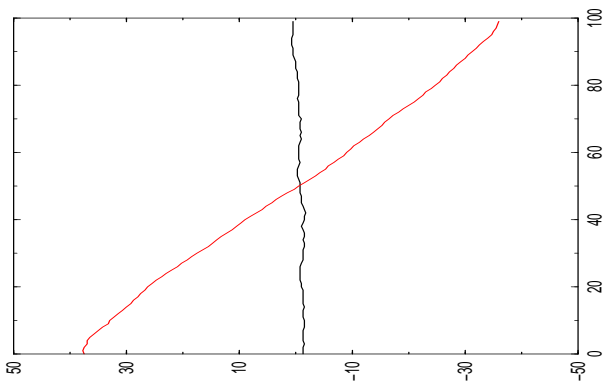}
\caption{\label{StrBelowTrans} Shape of the string: below the transition the charged
string is vertical. Above the transition it tilts away from the anisotropy axis.}
\end{figure}

\bigskip{}
\noindent We consider first the behavior of the order parameter \( \zeta  \)
as a function of \( \gamma _{num} \) for different sizes \( L \). A typical
result is displayed on figure \ref{GaussXiGamma}. Each point in this figure
corresponds to an independent Monte-Carlo simulation, i.e. given by generating
a new random configuration which is then thermalized by the simulated annealing
and brought to an effective zero temperature. Due to the double degeneracy of
the ground state it is the absolute value of \( \zeta  \) which has been plotted.
The behavior is the one expected by the analytical results. Further more the
critical coupling constant given by \( \gamma ^{c}_{Dnum} \) agrees with \( \gamma ^{c}_{D} \)
with an error of the order of \( 1\% \) as is given by fitting the numerical
data with (\ref{xiopt2}). Below the transition the mean angle is zero. A typical
shape of the string in this region is given in Figure \ref{StrBelowTrans}.
The string is vertical in average. Above the transition the mean angle increases
with \( \gamma _{num} \) in accordance with (\ref{xiopt2}). The string is
then tilted with respect to the main axis as shown also in Figure \ref{StrBelowTrans}.
No jump in the order parameter takes place which is an indication that the phase
transition is second order. The energy of the string has however been computed
as a function of \( \gamma _{num} \). It is continuous. The second derivative
is found on the other hand to diverge around the critical value.

\begin{figure}
\includegraphics[width=8cm,height=5cm]{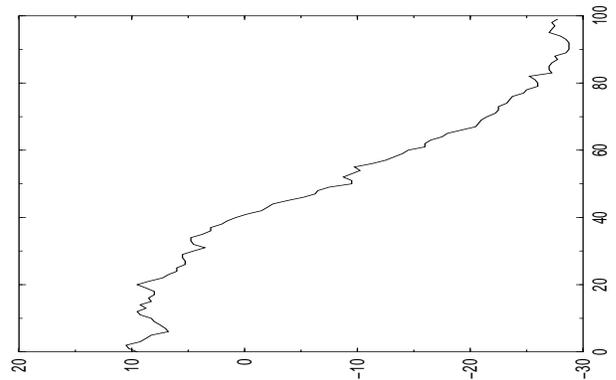}
\caption{\label{rstring}Rough inclined string at low temperature; orientational order
is preserved.}
\end{figure}

\bigskip{}
\noindent Next we take into account thermal fluctuations. When the temperature
is less than the activation energy of single angular kinks \( \Delta E_{1} \),
cf. (\ref{deltaE1}), the line is rough as shown by Figure \ref{rstring}. Above
\( \Delta E_{1} \) single angular kinks are present in addition to the roughening,
cf. Figure \ref{Figure 4}. The orientational order is then lost. These results
clearly confirm what has been said in the previous sections. At temperatures
higher than \( \Delta E_{1} \), thermal fluctuations average the tilt angle
to zero.

\subsection{Confinement regime}

The same approach is performed in the confinement regime.

\begin{figure}
\includegraphics[width=6.5cm,height=6cm,angle=-90]{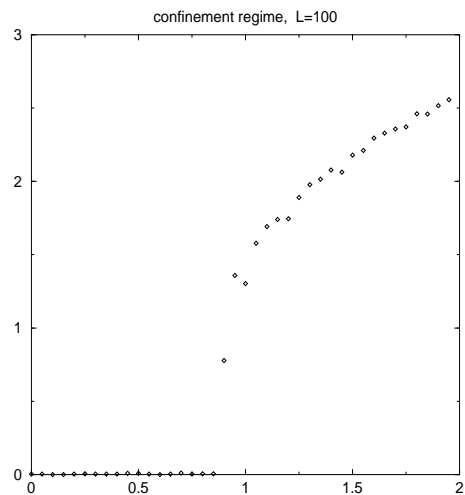}
\caption{\label{SOSXigamma}Absolute value of \protect\( \zeta \protect \) as a function
of \protect\( \gamma _{num}\protect \) in the confinement regime.}
\end{figure}

\begin{figure}
\includegraphics[width=6.5cm,height=6cm,angle=-90]{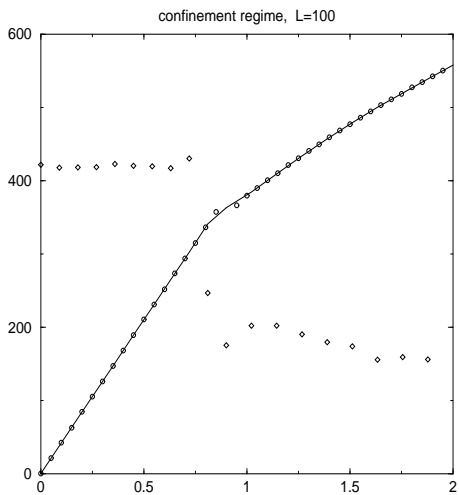}
\caption{\label{EnergyGamma}Energy of the string as a function of \protect\( \gamma _{num}\protect \).
The solid line corresponds to the energy of a string. A discontinuity is clearly
noticeable. The diamonds correspond to the first derivative of the latter.}
\end{figure}

\noindent Figure \ref{SOSXigamma} displays the absolute value of \( \zeta  \)
as a function of \( \gamma _{num} \). The behavior is the same as in the previous
case. The straight string is stable below the transition and tilts above. However
a jump in the order parameter is noticeable (especially in comparison with the
equivalent Figure \ref{GaussXiGamma} concerning the elastic regime), an indication
that the transition is first order. This is further confirmed by looking at
the energy displayed on Figure \ref{EnergyGamma}. It is clearly discontinues
as a function of \( \gamma _{num} \) at the transition. The numerical study
allows to capture the properties of the charged confined string in the whole
range of \( \gamma _{num} \). This is to be contrasted with the analytical
work where we could not consider the small tilt angle region. In contrast with
what has been said in \cite{teber} this regime is not manifested by an instability
of the string towards a modulation. The ground state given by the numerical
simulations is also the straight, either inclined or not, string.

\noindent At finite temperature the results for the confinement regime are qualitatively
the same as for the elastic one.

\section{Applications}
\label{applications}

\subsection{The solitonic lattice (\protect\( 2D\protect \))}

The present study should be connected, as has been said
in the Introduction, to the previous work on the statistical properties of solitons. Above,
we have been considering the domain-line as given. Actually, in quasi-one dimensional
systems, it has been shown in \cite{teber}, with the help of a mapping to the
ferromagnetic Ising model, that there are no strings, i.e. infinite domain lines
of solitons, at non-zero temperature. The latter emerge from a process of exponential
growth of aggregates of bisolitons below a crossover temperature \( T_{cro} \).
In the non-interacting case, \( T_{cro}=T_{0} \) and the aggregates are finite
rods perpendicular to the chains of the quasi-one dimensional system. At zero
temperature the rods cross the whole system forming the free domain lines of
solitons. In the interacting case the crossover temperature is lowered with
respect to \( T_{0} \) and becomes Coulomb dependent. In the region between
\( T_{0} \) and \( T_{cro} \), intrachain antiferromagnetic interactions,
within the ferromagnetic ground-state, take place in order for the bisolitons
to aggregate, cf. \cite{teber}. This is due to the important size-charging
energy of the aggregates. A crucial approximation is that we have been neglecting
shape instabilities in the interacting case. This reduced us to the region of
the phase diagram where \( \gamma _{D}<\gamma ^{c}_{D} \), cf. (\ref{gammaCriticGauss}).
At high coupling constants, \( \gamma _{D}\gg \gamma ^{c}_{D} \), strings at
\( T=0 \), or aggregates at finite temperatures should melt into a Wigner crystal
of solitons. At intermediate coupling constants the present theory, dealing
with shape instabilities should be relevant. It should be noted that in the
previous paper we have suggested that shape instabilities give rise to a modulated
string which would correspond to a single negative mode; as we have shown, the
instability at intermediate coupling constant, is towards a tilt of the interface
and affects a continuous set of modes.

\bigskip{}
\noindent We will first seek the stability of the string phase with respect
to the Wigner crystal of solitons at \( T=0 \). Then we will concentrate on
the non-zero temperature regime and look for the shape of the aggregates of
solitons which lead to strings. To make comparison with the previous paper easier,
notice that the present coupling constant, \( \gamma =\gamma _{D}/\log l_{D} \),
is similar to the coupling constant used in the previous paper \( \Gamma  \).
Otherwise we have tried to preserve the same notations.

\subsubsection{Zero temperature results: from strings to Wigner crystal of solitons}

We define \( \gamma ^{wc}_{D} \) as the coupling constant above which strings
melt into a Wigner crystal at \( T=0 \). We would like to check here that \( \gamma ^{wc}_{D}>\gamma ^{c}_{D} \)
in order to preserve the coherence of the arguments.

\bigskip{}
\noindent As the Coulomb interaction increases, the string of solitons tilts.
It's energy also increases. This leads to a dissociation of the string as soon
as it's energy per soliton \( \mu =\delta H^{(0)}/\delta N \) becomes larger
than the energy of the constituent soliton \( \mu _{sol} \). The latter corresponds
to \( 2J_{\perp }R/a_{x}-(ze)^{2}/\epsilon R \) where \( R\approx \sqrt{s/\nu } \)
is the radius of the Wigner Seitz cell including the soliton and a background
charge of density \( \nu /s \) with \( s=a_{x}a_{y} \). This energy can also
be written as
\[
\mu _{sol}=\frac{2J_{\perp }R}{a_{x}}\left[ 1-\nu \gamma \right] \]
where \( \gamma  \) is given by (\ref{gamma0}). The cohesion energy of the
string is thus
\begin{equation}
\label{Cohesion Energy}
E_{c}=J_{\perp }\left( \zeta +\frac{\gamma _{D}}{\sqrt{1+\zeta ^{2}}}\right) -\mu _{sol}.
\end{equation}
From (\ref{Cohesion Energy}) we see clearly that in the absence of Coulomb
interaction, which implies that \( \zeta =0 \), the vertical string is stable.
When the Coulomb interaction increases the tilt angle increases beyond the transition
to the tilted phase. Thus \( E_{c} \) increases from negative values. The melting
of the string is given by \( E_{c}=0 \) and, \( \zeta  \) being the optimal
angle of the string, (\ref{SOSsol}). Starting with \( \gamma _{D}\ll 1 \)
in \( E_{c}=0 \), (\ref{SOSsol}) leads to the boundary \( \gamma _{D}=1/\nu \gg 1 \),
in contradiction with the starting hypothesis. On the other hand, starting with
\( \gamma _{D}\approx 1/\nu  \) leads to the lowest boundary \( \gamma _{D}\approx \gamma _{D}^{*} \),
cf. (\ref{metast}), again in contradiction with the starting hypothesis. The
solution satisfying the self-consistent equation is thus in the intermediate
coupling constant range \( \gamma ^{*}_{D}<\gamma ^{wc}_{D}<1/\nu  \) and is
given by
\begin{equation}
\label{gammaDWC}
\gamma ^{wc}_{D}\approx 1/4\nu .
\end{equation}
We see from (\ref{gammaDWC}), that for a sufficiently diluted system, \( \gamma ^{wc}_{D}\gg \gamma ^{c}_{D} \),
cf. (\ref{critical1stOrder}). The tilted phase might thus be observed when
this condition is satisfied.

\subsubsection{Non-zero temperature results: shape of the aggregates}

\begin{figure}
\includegraphics[width=7cm,height=5cm]{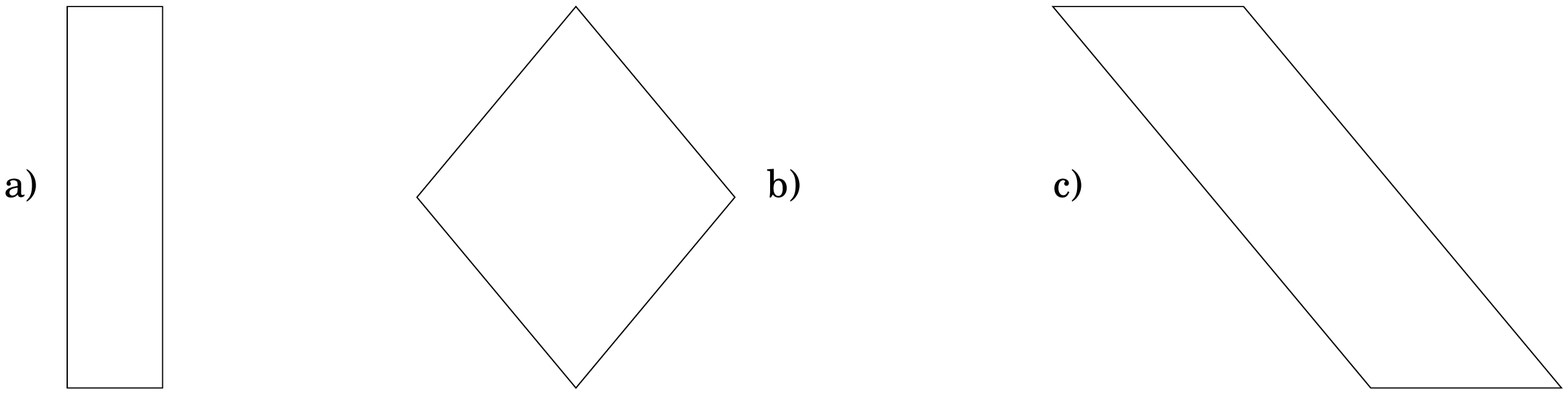}
\caption{\label{shapes}a) Rod in the absence of Coulomb interaction. The Coulomb energy
of the lozenge b) is lower than that of c). At finite temperature the aggregates
thus have the b) shape.}
\end{figure}

\noindent More exotic is the non zero temperature case. From the point of view
of interfaces, the latter roughen and thus collide at non-zero-temperatures.
Equivalently, from the point of view of solitons there are no infinite lines
but finite size aggregates. The size of the aggregates is approximately given
by the distance between two successive collision points of neighboring interfaces.
We address here the question of the shape of the charged aggregates of solitons
in the intermediate regime where \( \gamma ^{c}_{D}<\gamma _{D}\ll \gamma ^{wc}_{D} \).
As has been said above, in this regime, the aggregates must adapt their shapes
in order to minimize their electrostatic energy. An inclined aggregate, as in
Figure \ref{shapes} c), would thus be more favorable than the straight rod.
But at finite temperature angular kinks are present and they further reduce
the Coulomb energy. We can shown that a droplet with the lozenge shape of Figure
\ref{shapes} b) is favored in comparison with the previous aggregates. This
is done by evaluating the energies of these aggregates \( \Delta E_{b}=E_{b}-E^{(se)}_{b} \)
and \( \Delta E_{c}=E_{c}-E^{(se)}_{c} \) where their divergent self-energies
\( E^{(se)}_{b} \) and \( E^{(se)}_{c} \) have been subtracted. For the aggregate
c) the separation between the bikinks \( x_{opt} \) is taken as the equilibrium
separation under the combined action of the Coulomb force \( (ze)^{2}/\epsilon x^{2} \)
and the confinement one \( -J_{\perp }/a_{x} \), i.e. \( x_{opt}=\sqrt{s\gamma } \).
Moreover, the tilt angle corresponds to the optimal one. In the confinement
regime \( \gamma  \) is related to \( \zeta  \) with the help of (\ref{SOSsol}).
The regularized energies of the aggregates thus depend on one free parameter
\( \zeta  \). The later must be larger than \( 1/\sqrt{2} \) to be in the
stable confinement regime. We find that the difference between these energies
\( \Delta E_{c}-\Delta E_{b}>0 \) as a function of \( \zeta  \) so that the
energy of b) is always lower than that of c) above the critical coupling constant
\( \gamma ^{c}_{D} \).

\begin{figure}
\includegraphics[width=7.5cm,height=6cm]{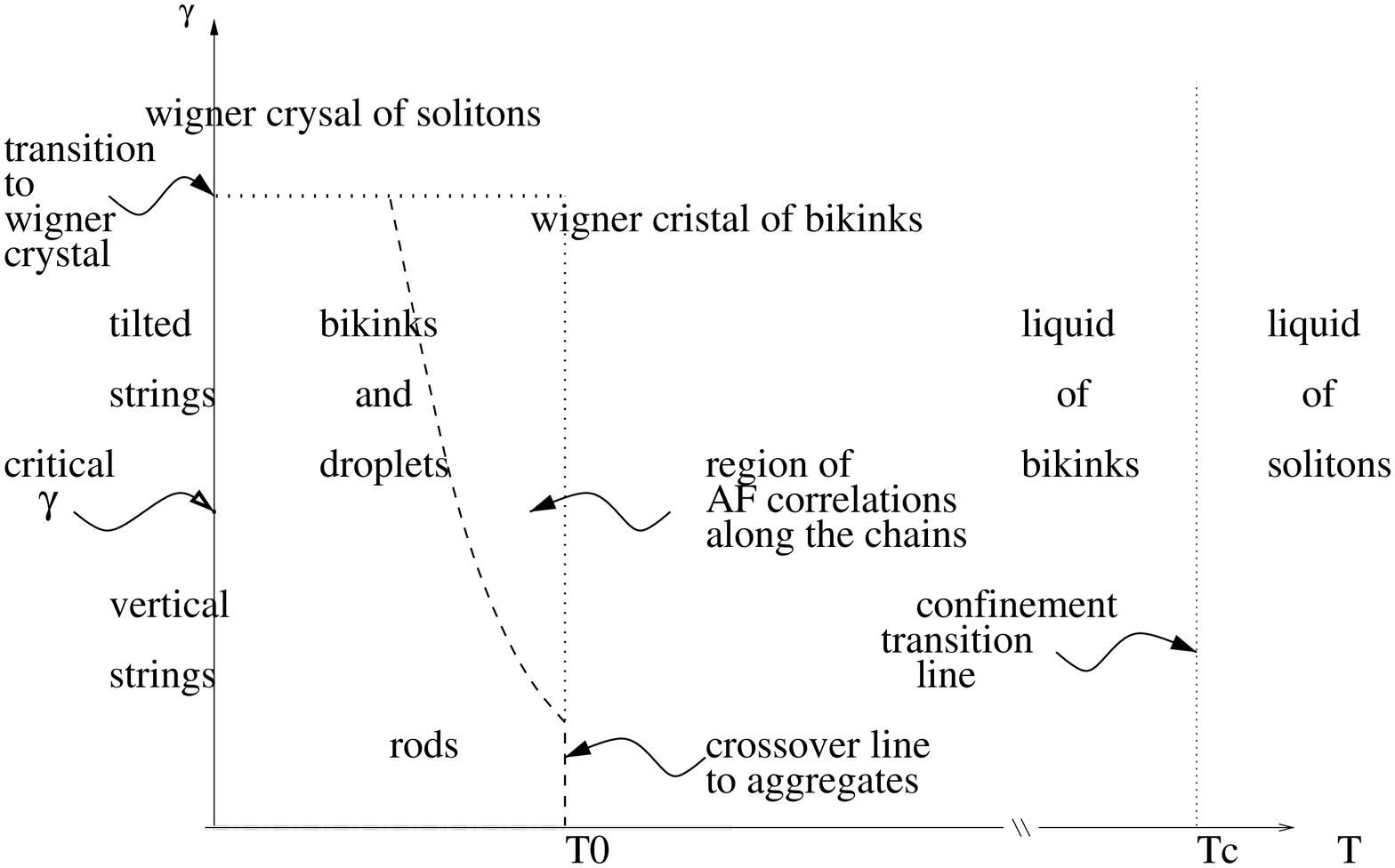}
\caption{\label{PhaseDiagr}Phase diagram of the \protect\( 2D\protect \) system as
a function of the coupling constant \protect\( \gamma _{0}\protect \) and temperature
\protect\( T\protect \). Details are given in the text.}
\end{figure}

\bigskip{}
\noindent The results obtained in this section are summarized on the phase diagram
Figure \ref{PhaseDiagr}.

\subsection{The solitonic lattice (\protect\( 3D\protect \))}

\label{3Dcase}We consider briefly the important case of three dimensional systems
where interfaces are domain walls. In \( 3D \) some rigorous statements allow
a good understanding of the uncharged case in the frame of quasi-one dimensional
systems, cf. \cite{bohr}. In the presence of Coulomb interactions we shall
thus follow \cite{bohr} as well as our present experience of the charged \( 2D \)
case.  Concerning notations, in the third direction, \( z \), the coupling
constant will be taken as \( J_{z} \) and the unit length as \( a_{z} \).

\subsubsection{Uncharged domain wall}

\noindent For the non-interacting case it has been shown in \cite{bohr} that
peculiarities are brought by the raise of dimensionality. In \( 3D \) the excitations
analogous to rods would be disc-like objects. However, rigorous statements concerning
the \( 3D \) case show that, at a temperature \( T_{3}\approx 8J_{\perp }/\ln (2/\nu )\ll T_{c}\approx J_{\perp }/\nu  \),
a density \( \nu _{c}=\nu -\nu _{n} \) of solitons condense into infinite anti-phase
domain walls perpendicular to the chains of the quasi-one dimensional system;
\( \nu _{n} \) then corresponds to the density of finite size solitons below
\( T_{3} \). Therefore, the crossover regime of growing rods, which took place
below \( T_{0} \) in \( 2D \), cannot take place in \( 3D \) and is replaced
by the transition at \( T_{3} \).

\noindent The effect of thermal fluctuations on the walls is well known. Several
authors have shown (see references in \cite{forgacs}) that, in \( 3D \) the
hight-hight correlation function remains finite at \( T<T_{R} \) whereas it
diverges logarithmically for \( T\geq T_{R} \), where \( T_{R}>0 \) is the
roughening temperature. From these considerations \( T_{3} \) may be considered
as the roughening temperature. The existence of walls in \( 3D \) at finite
temperatures is compatible with the fact that in Ising-like systems there is
\( 2D \) long-range order at low temperatures.

\subsubsection{Charged domain wall versus Wigner crystal}

\noindent The three-dimensional case of charged walls can be treaded the same
way as the two-dimensional case. We will therefore consider again the elastic
and confinement regimes. The Coulomb interaction will lead to a tilting of the
wall here too. Two tilt angles can be defined now, because the anisotropy axis
becomes an anisotropy plane. We will thus consider \( \zeta _{y} \) which corresponds
to the previous, in plane, tilt angle and \( \zeta _{z} \) which is the tilt
angle with respect to the third direction. The mean-field hamiltonian of the
system is then
\begin{widetext}
\begin{equation}
\label{H3D}
H_{p}/J_{y}=\zeta _{y}^{p_{y}}+\alpha _{\perp }\zeta ^{p_{z}}_{z}+\gamma _{D}\left[ \frac{1}{\sqrt{1+\zeta ^{2}_{z}}}\left( 1-\frac{1}{2}\log \left( \frac{1+\zeta ^{2}_{y}}{1+\zeta ^{2}_{z}}\right) \right) -1\right]
\end{equation}
\end{widetext}
where, \( \alpha _{\perp }=J_{z}/J_{y} \), \( \gamma _{D}=e^{2}l_{D}/\epsilon s_{\perp }J_{y} \),
\( s_{\perp }=a_{y}a_{z} \) and in the last term, the contribution of the straight
domain wall has been substracted. Notice that it is only in the case where \( \zeta _{y}=\zeta _{z}=\zeta  \)
that we gain a global factor \( 1/\sqrt{1+\zeta ^{2}} \) as in the \( 2D \)
case. The logarithm is therefore the signature of a non-homogeneous charge dilatation
within the wall.

\bigskip{}
\noindent Our aim is to determine the phase diagram, \( \gamma _{D} \) versus
\( \alpha _{\perp } \) in both regimes, i.e. \( p_{y,z}=1,2 \). Among the
four possible cases we will consider only the full elastic (\( p_{y}=p_{z}=2 \))
and mixed (\( p_{y}=2 \), \( p_{z}=1 \)) regimes.

\begin{figure}
\includegraphics[width=7.5cm,height=6cm]{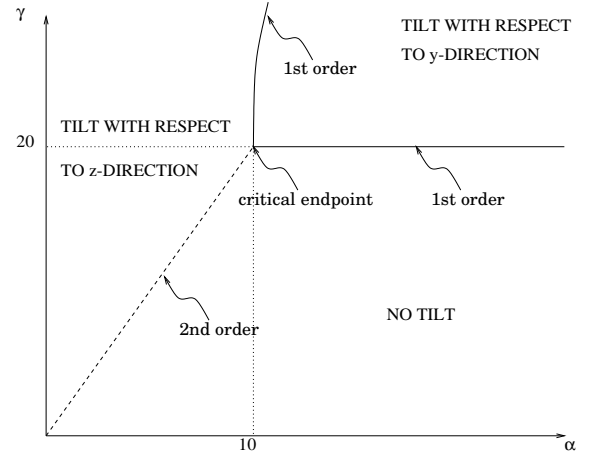}
\caption{\label{wall elastic} Phase diagram of the elastic wall: \protect\( \gamma _{D}\protect \)
as a function of \protect\( \alpha _{\perp }\protect \).}
\end{figure}

\bigskip{}
\noindent In the elastic regime, \( p_{y}=p_{z}=2 \), the phase diagram is
displayed on Figure \ref{wall elastic}. For \( \gamma _{D}>2\alpha _{\perp } \),
the stable solution is given by
\begin{equation}
\label{phase00}
\zeta _{y}=\zeta _{z}=0
\end{equation}
In the region \( \gamma _{D}<2\alpha _{\perp } \), the stable solution corresponds
to
\begin{equation}
\label{phase0z}
\zeta _{y}=0\qquad \zeta _{z}=\sqrt{\frac{\gamma _{D}}{2\alpha _{\perp }}-1}
\end{equation}
For \( \gamma ^{s}_{D}\approx 16<\gamma _{D}<(2\alpha _{\perp })^{3/2}\sqrt{\log \alpha _{\perp }/4} \)
a third phase, corresponding to
\begin{equation}
\label{phasey0}
\zeta _{y}\approx \left( \frac{\gamma _{D}}{6}\log \left( \frac{\gamma _{D}}{4}\right) \right) ^{1/3}\qquad \zeta _{z}=0
\end{equation}
is stable. The domain of validity of this solution overlaps the ones of the
two previous solutions. With the help of energetic considerations, we find that
a line of first order critical points, at \( \gamma _{D}\approx 20 \), separates
phase (\ref{phase00}) from phase (\ref{phasey0}). By the same way, a line
of first order critical points has been found to separate phase (\pageref{phase0z})
from phase (\ref{phasey0}). The line \( \gamma _{D}=2\alpha _{\perp } \) is
a line of second order critical points which terminates at the critical end-point
\( \alpha _{\perp }=10 \), \( \gamma _{D}=20 \). Phases which would have both
angles none-zero appeared to be unstable.

\bigskip{}
\noindent A similar analysis can be done for quasi-two dimensional (\( Q2D \))
systems where the interface is elastic within the planes, i.e. \( p_{y}=2 \)
and in the confinement regime between planes, i.e. \( p_{z}=1 \). For \( Q2D \)
systems, the coupling in the third direction is weak with respect to the coupling
in other directions, i.e. \( \alpha _{\perp }<1 \). In the region of the phase
diagram where \( \alpha _{\perp }<1 \) the solution is thus given by
\begin{equation}
\label{ConfintAngle}
\zeta _{y}=0\qquad \zeta _{z}=\frac{\gamma _{D}}{2\alpha _{\perp }}+\sqrt{\left( \frac{\gamma _{D}}{2\alpha _{\perp }}\right) ^{2}-1}
\end{equation}
and is stable for \( \gamma _{D}>2\alpha _{\perp } \). As before, when \( \gamma _{D}<2\alpha _{\perp } \),
the untilted solution is stable for small \( \alpha _{\perp } \) and is separated
from the previous by a line of second order critical points. From (\ref{phase0z})
and (\ref{ConfintAngle}), we see that for a given value of the ratio \( \gamma _{D}/2\alpha _{\perp } \),
the tilt is stronger in the mixed regime than in the pure elastic regime. This
can be interpreted by considering the equilibrium position of a constituent
of the wall due to the constant confinement force and the constant electric
force due to the charged wall; a priori, the equilibrium is reached only at
\( \gamma _{D}=\alpha _{\perp } \), a feature which is peculiar to the confinement
regime. This implies that the wall should melt as soon as \( \gamma _{D}>\alpha _{\perp } \)
\textit{or,} if the Coulomb energy scale is sufficiently weak, that the wall
should strongly incline to reduce the coupling constant. Allowing the wall to
tilt, the equilibrium equation, \( \gamma _{D}=\alpha _{\perp } \), is generalized
to
\[
\frac{\gamma _{D}}{\alpha _{\perp }}\left( 1-\frac{1}{2}\log \left( 1+\zeta ^{2}_{z}\right) \right) =1.\]
which leads to the following minimal stability angle
\[
\zeta _{z}=\exp \left( 2\left( 1-\frac{\alpha _{\perp }}{\gamma _{D}}\right) \right) -1.\]
above which the wall stabilizes at the equilibrium angle (\ref{ConfintAngle}).

\bigskip{}
\noindent In \( 3D \), the situation is thus richer than in the \( 2D \) case
with the appearance of a tricritical point and quantitative differences between
the various regimes but, qualitatively, the same phase diagram for all of them.
In particular, the tilting of the interface takes place with respect to a single
anisotropy axis and not both. In quasi-two dimensional systems, the wall should
tilt with respect to the third direction rather than \( y- \)direction and
the tilting is stronger in the confinement regime to prevent the melting. Another
difference with the \( 2D \) case is that thermal fluctuations do not affect
these results, as long as we are below the roughening temperature of the wall.

\subsection{Stripes in oxides}

Up to know the whole theory is model independent. As we have already mentioned
our results could be applied to doubly commensurate charge-density waves such
as polyacetylene. We will however consider physical systems which are quite
extensively studied nowadays. These are the oxides, e.g. cuprate, nickelate,
manganese and the new ferroelectric state. The latter will be considered in
the next section. We will focus here on the phenomenology of stripes with the
help of the above theory. We concentrate only on the long distance properties,
namely, the fact that stripes consist of lines or walls of holons which are
similar to the charged solitons we have been considering. The latter are thus
charged interfaces as the ones described in this paper.

\bigskip{}
\noindent The oxides we consider are quasi-two dimensional materials formed
by coupled planes. At intermediate values of doping, a \( 3D \) ordering of
the stripes might have been seen in \( La_{1.6-x}Nd_{0.4}Sr_{x}CuO_{4} \) and
\( La_{2-x}Sr_{x}NiO_{4+\delta } \), cf. \cite{Tranq} and references therein.
The impact of the pinning by the lattice and the Coulomb interactions play a
crucial role in these systems.

\noindent In \( La_{1.6-x}Nd_{0.4}Sr_{x}CuO_{4} \) neutron diffraction studies
have revealed diffuse magnetic peaks along the \( z \)-direction, in the stripe
phase, which is consistent with stripes as one-dimensional interfaces. Moreover,
\( X \)-ray studies reveal that stripes are parallel in next-nearest-neighboring
layers but orthogonal between nearest layers. This \( 3D \) ordering is probably
due to a pinning of the stripes by the lattice structure.

\noindent The case of nickelates is more interesting. In \( La_{2-x}Sr_{x}NiO_{4+\delta } \)
there is a body-centered stacking of parallel elastic stripes with only weak
perturbations due to the lattice. This \( 3D \) ordering is due to Coulomb
repulsion and might correspond to a domain wall with a step-like profile suggesting
a confinement regime in the third direction. This justifies the mixed confinement-elastic
regime we have considered in \ref{3Dcase}. Experimental results do not however
exclude the interpretation of this \( 3D \) ordering as a Wigner crystal of
lines. The latter hypothesis allows us to focus on strings within planes of
the quasi-two dimensional system, as we are going to do in the next paragraph.

\bigskip{}
\noindent Following Wakimoto \& al \cite{incli}, we are going to recall briefly
here some experimental results related to the tilting of string-like stripes
in \( La_{2-x}Sr_{x}CuO_{4} \).

\begin{figure}
\includegraphics[width=7cm,height=7cm]{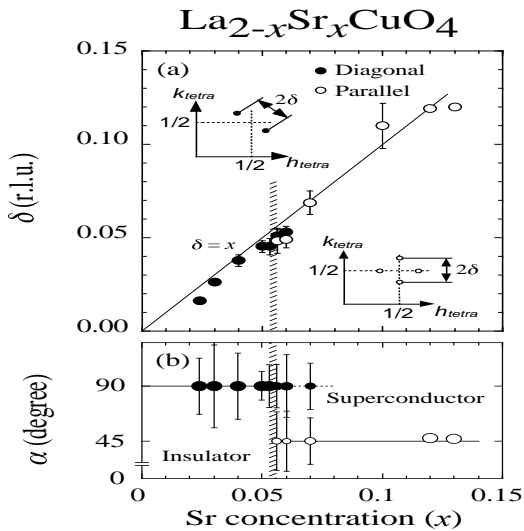}
\caption{\label{cuprate}Tilting of stripes in a cuprate oxide (from \cite{incli}).}
\end{figure}

\noindent The experimental results are summarized on Figure \ref{cuprate} which
has been taken from \cite{incli}. In Figure \ref{cuprate} (a), \( \delta =2\pi /d \)
where \( d \) is the distance between stripes. At low doping, the inset on
the left of this figure, shows magnetic Bragg peaks corresponding to a one-dimensional
spin modulation which is inclined with respect to the reference tetragonal axis.
At higher doping, the inset on the left, shows that this spin modulation, along
with the charge stripes, have rotated by an angle of \( 45 \) degrees. The
main figure, together with Figure \ref{cuprate} (b) which displays the tilt
angle as a function of the concentration of dopant, show that this shape instability
takes place around \( x_{c}\approx 0.05 \) where \( \delta  \) increases linearly
with \( x \). Moreover, the authors report a weak dependence of the magnetic
peaks on the third direction. This implies that the spin modulations are weakly
correlated between successive \( CuO_{2} \) layers. Thus, the stripes can be
considered with a good accuracy as charged strings which justifies our starting
hypothesis.

\noindent The stripe instability observed experimentally might be due to the
competition between the long-range Coulomb interaction and an anisotropic interaction.
To relate this instability to an increase of the concentration of dopants \( x \)
we should probably take into account structural changes. As argued by Tranquada
\& al \cite{Tranq}, the lines of charge might be pinned along the direction
of the tilt of the \( CuO_{6} \) octahedron. This tilt direction varies with
doping. It is along the orthorhombic axis at low doping and along the tetragonal
axis at higher doping. These axis are \( 45 \) degrees from each-other. At
an intermediate doping, corresponding to the critical value \( x_{c} \), a
\( 45 \) degrees tilt of the stripes, satisfying the structural constraints,
can take place.

\noindent Simple considerations related to the competing energy scales are in
favor of such an order of magnitude for the tilt angle. First, in this system
\( e^{2}/\epsilon a\sim 0.1-1eV \) where \( a \) is a unit length of the tetragonal
structure. Second the exchange energy \( J_{\perp } \) is of the order of \( 0.13eV \).
Thus \( \gamma \sim 0.4-4.5 \) where (\ref{gamma0}) has been used. If we consider
that the Debye length is of the order of the inter-stripe distance \( 1/\delta  \),
at the transition \( x_{c} \) we have \( l_{D}\sim 20a \). Thus \( \gamma _{D}=\gamma \ln (l_{D}/a)\sim 1.2-13.5 \).
Using (\ref{xiopt2}), the maximum tilt angle compatible with these values is
around \( 55 \) degrees. A tilt angle of \( 45 \) degrees can be reached with
a \( \gamma _{D}\sim 4.0 \) which is clearly in the range of allowed values
above. Hence, the competing interactions can be responsible for such a tilt.

\noindent This tilt could reduce the electrostatic energy of the stripe as can
be seen by the following argument. At low doping, i.e. in the diagonal stripe
phase, the hole concentration along the stripe is \( 0.7 \) hole/Cu. This concentration
is reduced to \( 0.5 \) hole/Cu in the collinear stripe phase, cf. Wakimoto
\& al. The density of charges along the stripe is thus reduced by \( \sqrt{1+\zeta ^{2}} \),
with \( \zeta =1 \) for the present \( 45 \) degrees tilt angle. This dilatation
factor reduces the electrostatic energy of the string as has already been seen
in (\ref{MFfreeEn}). This could hold assuming that the number of holes\( / \)stripe
is constant. The added dopants would lead to the formation of new stripes. The
latter become tightly packed at high doping as suggested by the linear relation
\( \delta \approx x \).

\subsection{Uniaxial ferroelectrics }

Recently a ferroelectric Mott-Hubbard state has been observed in quasi 1D organic
superconductors such as \( (TMTTF)_{2}X \). We give a brief account on the
origin of this phase and connect the results to our present study. Experimental
and theoretical results are given in detail in \cite{Monceau}.

\noindent In the system mentioned above \( TMTTF \) are molecules composing
the stacks and therefore giving rise to the one dimensional nature of the compound.
\( X \) are ions placed in the vicinity of the molecules. At large temperatures,
of the order of \( 100K \), the ions reorder through a uniform shift which
gives rise to a ferroelectric phase. The latter has been detected via the measurement
of the dielectric susceptibility which shows a gigantic anomaly at low frequency
around \( T_{fe}\sim 150K \). At low temperature domain lines, separating domains
with opposite polarization, should be observed. Their existence is necessary
to minimize the external electric field generated by charges accumulating at
the boundary of the sample. Moreover, as the interface corresponds to a jump
in the polarization it is charged. Following \cite{Monceau} we shall define
as \( \alpha - \)solitons the elementary constituents of the interface which
connect domains with opposite polarization. They carry a fractional charge \( ze \)
with \( z=2\alpha /\pi  \).

\noindent Two facts suggest that this system is a good candidate for the observation
of charged interfaces. First the background dielectric susceptibility \( \epsilon _{0}\sim (\omega _{p}/\Delta )^{2}\sim 10^{4} \)
is large. Further more \( \pi - \)solitons are also present and they screen
the Coulomb interaction. This weakens the Coulomb interaction. In such systems,
the strings of \( \alpha - \)solitons should thus be stable against a melting
towards a Wigner crystal. Bubble-like aggregates of \( \alpha - \)solitons,
as the ones described in \ref{applications}, could also be observed at higher
temperatures.

\section{Conclusion}

The statistical properties of charged interfaces, strings and walls, has been studied. We
have found that shape instabilities, due to competing interactions, play a fundamental
role. They manifest themselves through zero-temperature phase transitions from
which new ground-states emerge where the interface is tilted. The \( 2D \)
case of the string has been extensively studied. For small tilt angles the string
is elastic and the transition to the tilted ground state is second order. On
the other hand a first order transition takes place if we are beyond the limit
of validity of the elastic regime, i.e. in the confinement regime. At non-zero
temperatures it has been shown that in either regimes a pure Coulomb roughening
of the string, due to the proliferation of angular-bikinks, exists in addition
to the usual roughening as soon as \( T>0 \). At higher temperatures angular-kinks
are thermally activated and connect the degenerate ground-states. These results
have been confirmed with the help of a numerical approach based on the Monte
Carlo - Metropolis algorithm. We have then related the present study to the
general problem of the statistical properties of solitons in charge-density
wave systems in either \( 2 \) or \( 3 \) dimensions. In \( 2D \) systems
the string has been shown to emerge from lozenge-like aggregates of solitons
upon decreasing temperature. In \( 3D \), the additional coupling constant
in the third direction enriches the tilted phases with the appearance of a tricritical
point separating them. In the confinement regime, a disintegration of the wall
is favored but the solitons might order in the form of a strongly inclined wall.
Applications concerning stripes in oxides and a possible explanation of their
tilt, as well as charged strings in uniaxial ferroelectric experiments, have
also been considered.

\section*{Aknowledgments}

\noindent I would like to thank S. Brazovskii for helpful and inspiring discussions
from the very beginning of this work. I would like to thank A. Bishop for discussions
and his hospitality at the Theoretical Division and Center of Non-Linear Studies,
of Los Alamos.

\section*{Appendix A}

\noindent In this appendix we give some details on the screening in the ensemble
of solitons. Screening makes the energy extensive and gives a meaning to the
thermodynamic limit as has been said all along the paper. In the following we
will work in the RPA approximation which is very well known in the field on
many-body theory. This allows us to derive the Debye length and the electrostatic
potential in two cases: two and three-dimensional screening.

\noindent We consider first a single plane embedded in \( 3D \) space. This
case corresponds to two-dimensional screening. Supposing the Coulomb field \( \phi  \)
slowly varying, it obeys the following semi-classical equation
\begin{equation}
\label{General Poisson}
-\epsilon \Delta \phi (\overrightarrow{R})=-4\pi e^{2}(\Pi ^{(0)}*\phi )(\overrightarrow{R})\delta (z)+4\pi e^{2}\rho _{0}(\overrightarrow{r})\delta (z)
\end{equation}
where \( \overrightarrow{R} \) is a \( 3D \) vector associated with the \( 3D \)
Coulomb field, \( \overrightarrow{r} \) is a \( 2D \) vector indexing the
charges which are embedded in the \( z=0 \) plane and \( \rho _{0} \) is some
external charge. \( \Pi ^{(0)}(x,y) \) is the response function, or polarization
part, of the soliton system which is related to the correlation function by
fluctuation dissipation theorem
\begin{equation}
\label{tPigeneral}
T\Pi ^{(0)}(\overrightarrow{r})=<\delta \rho (\overrightarrow{r})\delta \rho (\overrightarrow{0})>
\end{equation}
where \( \delta \rho (\overrightarrow{r})=\rho (\overrightarrow{r})-\nu  \),
\( \rho (\overrightarrow{r}) \) being the density of solitons and \( \nu  \)
the background neutralizing charge.

\noindent The first term on the right of (\ref{General Poisson}) corresponds
to the screening charge. This term implies linear screening and dispersion via
the convolution. The Coulomb field is then
\begin{equation}
\label{CoulProp}
\phi (\overrightarrow{k},z=0)=\frac{2\pi e^{2}\rho _{0}(\overrightarrow{k})}{\epsilon |k|+2\pi e^{2}\Pi ^{(0)}(\overrightarrow{k},z=0)}
\end{equation}
where \( \overrightarrow{k} \) is the reciprocal vector associated with \( \overrightarrow{r} \).
The fact that the potential is evaluated at \( z=0 \) implies that the field
lines are confined to the same space as the particles, cf. \cite{Cornu}. Expression
(\ref{CoulProp}) leads to the Debye screening of the field with a Debye length
\begin{equation}
\label{2DDebye}
\frac{1}{l_{2D}}=\frac{2\pi e^{2}}{\epsilon }\Pi ^{(0)}(k=0).
\end{equation}
Also the \( k- \)term in the denominator of (\ref{CoulProp}) is a feature
of \( 2D- \)systems. In real space this leads to an algebraic screening instead
of the exponential one, cf. \cite{Cornu} for detailed studies of screening
effects in Coulomb fluids and below for an example.

\noindent We consider now the three dimensional case of independent planes embedded
in a neutral media. Such a system of decoupled planes brings a \( 3D- \)screening.
The screening charge is then a function of the three-dimensional vector \( \overrightarrow{R} \).
But, as the planes are independent, the bare correlation function corresponds
to \( \Pi ^{(0)}(\overrightarrow{R})=\Pi ^{(0)}(\overrightarrow{r})\delta (z) \).
By the same arguments as in the first section we can compute the coulomb field
which reads
\begin{equation}
\label{3DCoulomb}
\phi _{3D}(\overrightarrow{K})=\frac{4\pi e^{2}\rho _{0}(\overrightarrow{K})}{\epsilon K^{2}+4\pi e^{2}\Pi ^{(0)}(\overrightarrow{k},z=0)/a_{z}}
\end{equation}
where we consider only the \( z=0 \) plane subject to the bulk screening. The
three-dimensional Debye length is then given by
\begin{equation}
\label{3DDebye}
\frac{1}{l^{2}_{3D}}=\frac{4\pi e^{2}}{\epsilon a_{z}}\Pi ^{(0)}(k=0)
\end{equation}
where \( a_{z} \) is the inter-plane distance. In real space (\ref{3DCoulomb})
leads to the usual exponential screening.

\noindent Finally we derive the polarization part for the present problem. This
gives explicit expressions for the screening lengths and for \( \phi (\overrightarrow{r}) \).
Below the crossover transition \( T_{0} \)
\begin{widetext}
\begin{equation}
\label{G0(x,y)}
T\Pi ^{(0)}(x,y)\approx \frac{1}{2s^{2}}\nu a_{x}\exp \left( -\frac{2y}{l_{\perp }}\right) \left( 2\delta (x)+\delta (x+a_{x})+\delta (x-a_{x})\right)
\end{equation}
\end{widetext}
corresponding to the aggregation of the solitons into rods of length \( l_{\perp } \),
cf. \cite{bohr}. Fourier transforming (\ref{G0(x,y)}) we get
\[
T\Pi ^{(0)}(\overrightarrow{k})\approx \frac{\nu }{4sa_{y}}(1+\cos (k_{x}a_{x}))\frac{l_{\perp }}{1+k^{2}_{y}l^{2}_{\perp }/4}.\]
Thus \( \Pi ^{(0)}(\overrightarrow{k})\approx \frac{\nu l_{\perp }}{2sa_{y}T} \)
and, as we work at constant density of solitons \( \nu  \), \( l_{\perp }\approx \sqrt{2\nu }\exp (2J_{\perp }/T) \).
With the help of (\ref{2DDebye}) and (\ref{3DDebye}), this yields
\begin{eqnarray}
l_{2D}=\frac{2\epsilon sa_{y}T}{\pi e^{2}(2\nu )^{3/2}}\exp \left( -\frac{2J_{\perp }}{T}\right),
\nonumber\\
l_{3D}=\sqrt{\frac{\epsilon sa_{z}T}{\pi e^{2}(2\nu )^{3/2}}}\exp \left( -\frac{J_{\perp }}{T}\right) .
\label{DebLengths}
\end{eqnarray}

\noindent In the \( 2D- \)screening regime, the field can then be written,
in the long distance limit with respect to the coarse graining length \( a_{x} \)
\begin{equation}
\label{Pot2D}
\phi (\overrightarrow{k})=\frac{2\pi e^{2}}{\epsilon }\frac{1+k^{2}_{y}l^{2}_{\perp }/4}{|k|(1+k^{2}_{y}l^{2}_{\perp }/4)+1/l_{D}}
\end{equation}
which reflects the anisotropy of the problem. To give an idea of the algebraic
screening we Fourier transform (\ref{Pot2D}) in the limit \( k_{y}l_{\perp }\rightarrow 0 \),
i.e. far enough from the rod. This yields
\[
\phi (\overrightarrow{r})\approx \frac{2\pi e^{2}}{\epsilon }\frac{l_{D}^{2}}{r^{3}}\]
in the limit \( r\gg l_{D} \). We see that the algebraic tail leads to a slow
decay of the potential in contrast to the usual exponential one. However this
type of interaction, of the dipolar-type, beyond the screening length are compatible
with the growth of the aggregates, cf. \cite{teber}. For more details on the
peculiar screening properties of such system with anisotropic intrinsic charges
see \cite{these}.

\noindent \section*{Appendix B}

\noindent Our purpose is mainly to show the differences and common points between
the two interface models. The approach is based on the transfer matrix technic.
In the present case, this formulation cannot be used straight forwardly because
of the long-range interactions in (\ref{TotHamExplicit}). Some results obtained
in the non-interacting case should however remain in the presence of the long
range interaction. It is on such a property that we focus now.

\noindent The associated Hamiltonian of the problem is
\begin{equation}
\label{HforTM}
\beta H_{p}=\frac{K}{a^{p-1}}\sum ^{L-1}_{y=0}|x_{y+1}-x_{y}|^{p}
\end{equation}
where the short distance cut off has been taken explicitly into account. The
transfer matrix \( \widehat{T_{a}} \) is defined as an integral operator with
eigenfunctions \( \psi (x) \) whose kernel is expressed in terms of (\ref{HforTM}).
Namely
\begin{equation}
\label{IntEqForTa}
\widehat{T_{a}}\psi (x)=\int dx'\exp \left( -\frac{K}{a^{p-1}}|x-x'|^{p}\right) \psi (x').
\end{equation}

\noindent The free problem is simple enough to be treated just with the integral
equation (\ref{IntEqForTa}). For a more exhaustive account on the general results
one can obtain with help of the transfer matrix technic see \cite{suris}.

\noindent For both models the eigenfunctions of the transfer matrix are simply
plane waves
\[
\psi _{k}(x)\approx \exp (ikx).\]
This amounts to Fourier transform the kernel of (\ref{IntEqForTa}). The eigenvalues
of the transfer matrix associated to both interface models are thus
\begin{equation}
\label{EkSOS}
\epsilon ^{G}_{k}\approx \frac{k^{2}}{2K}\qquad \qquad \epsilon ^{SOS}_{k}\approx \log \left( K^{2}+k^{2}\right) .
\end{equation}
By comparing the two spectrums in (\ref{EkSOS}) we see clearly that both models
have the same long distance, i.e. \( k\rightarrow 0 \), properties.

\end{document}